\documentclass[aps,prd,preprint,showpacs,superscriptaddress,tightenlines]{revtex4}
\usepackage{graphics}
\usepackage{tabularx}

\begin{document}
\title{Effects of quark family nonuniversality in $SU(3)_c\otimes SU(4)_L\otimes U(1)_X$ models}
\author{Jorge L. Nisperuza}
\affiliation{Escuela de F\'\i sica, Universidad Nacional de Colombia,
A.A. 3840, Medell\'\i n, Colombia}
\author{Luis A. S\'anchez}
\affiliation{Escuela de F\'\i sica, Universidad Nacional de Colombia,
A.A. 3840, Medell\'\i n, Colombia}


\begin{abstract}
Flavor changing neutral currents arise in the $SU(3)_c\otimes SU(4)_L\otimes U(1)_X$ extension of the standard model because anomaly cancellation among the fermion families requires one generation of quarks to transform differently from the other two under the gauge group. In the weak basis the distinction between quark families is meaningless. However, in the mass eigenstates basis, the Cabibbo-Kobayashi-Maskawa mixing matrix motivates us to classify left-handed quarks in families. In this sense there are, in principle, three different assignments of quark weak eigenstates into mass eigenstates. In this work, by using measurements at the Z-pole, atomic parity violation data and experimental input from neutral meson mixing, we examine two different models without exotic electric charges based on the 3-4-1 symmetry, and address the effects of quark family nonuniversality on the bounds on the mixing angle between two of the neutral currents present in the models and on the mass scales $M_{Z_2}$ and $M_{Z_3}$ of the new neutral gauge bosons predicted by the theory. The heaviest family of quarks must transform differently in order to keep lower bounds on $M_{Z_2}$ and $M_{Z_3}$ as low as possible without violating experimental constraints.
\end{abstract}

\maketitle

\section{\label{sec:intr}Introduction}
Two of the most mysterious aspects in modern particle physics are the masses and mixings of the elementary fermions and the number of fermion generations in nature: the flavour problem. The standard model (SM), in which each family is anomalyfree by itself, does not fix the number of generations except by the indirect bound coming from the asymptotic freedom of QCD according to which this number must be less than 9. On the other hand it is known, on the theoretical side, that the flavor democracy approach \cite{fd} and some mechanisms of dynamical electroweak symmetry breaking \cite{holdom} favor the existence of a fourth family, and, on the experimental side, that precision electroweak data do not totally exclude it \cite{4f}. Moreover, two and even three additional generations could also be allowed \cite{56f}. However, in the SM, the LEP data on invisible Z boson decay show that there are three SM families with light neutrinos \cite{pdg}.

Two alternative scenarios, which provide some insight for the solution of this puzzle by relating the number of generations to the cancellation of chiral anomalies, have been proposed in the literature. In one of them anomalies constrain the number of generations provided their cancellation takes place either in a nonsupersymmetric $SU(3)_c\otimes SU(2)_L\otimes U(1)_Y$ theory that lives in a six-dimensional spacetime \cite{dobrescu}, or in a sixdimensional
($1,1$) supersymmetric gauge theory \cite{WaYa}. In the other one the SM is extended either to the gauge group $SU(3)_c\otimes SU(3)_L\otimes U(1)_Y$ (the 3-3-1 model) \cite{su3,331} or to the gauge symmetry $SU(3)_c\otimes SU(4)_L\otimes U(1)_Y$ (the 3-4-1 model) \cite{su41,su42,sap}, with anomalies cancelling among the families (three-family models) and not family by family as in the SM. In the 3-3-1 extension this happens only if we have an equal number of left-handed triplets and antitriplets, taking into account the color degree of freedom. Correspondingly, an equal number of 4-plets and $4^*$-plets is required in the 3-4-1 extension. As a consequence, the number of fermion families $N_f$ must be divisible by the number of colors $N_c$ of $SU(3)_c$, being $N_f=N_c=3$ the simplest solution. 

One additional motivation to study the second scenario comes from the fact that it has been recently recognized as the simplest SM extension required for the implementation of the little Higgs mechanism \cite{little1}. Even though we will not be concerned here with this alternative proposal to solve the so-called hierarchy problem, we notice that in the simplest little Higgs scenario the SM gauge group is enlarged to $SU(3)_L\otimes U(1)_X$. This model, however, lacks a quartic Higgs coupling which can be generated in a $SU(4)_L\otimes U(1)_X$ extension \cite{little1}. The complete anomaly-free fermion sector for these two little Higgs models has been studied in detail in Ref.~\cite{little2}, with direct generalization to $SU(N)_L\otimes U(1)_X$ with $N>4$ but, at present, there is not motivation to go beyond $N=4$. Conspicuously, in the little Higgs scenario both the 3-3-1 extension and the 3-4-1 one are three-family models in which all the exotic fermion fields have only ordinary electric charges.

In this paper we will be involved with the 3-4-1 extension of the SM. In this regard, a recent systematic analysis has shown that, by restricting the fermion field representations to particles without exotic electric charges (that is, electric charges different from $\pm 2/3$ and $\pm 1/3$ for exotic quarks and different from 0 and $\pm 1$ for exotic leptons) and by paying due attention to anomaly cancellation, a few different models are obtained; while by relaxing the condition of nonexistence of exotic electric charges, an infinite number of models can be generated \cite{sap}. The restriction to ordinary electric charges in the fermion, gauge boson and scalar sectors, allows only for two different possibilities for the simultaneous values of the parameters $b$ and $c$ in the most general expression for the electric charge generator in $SU(4)_L\otimes U(1)_X$
\begin{equation}\label{ch}
Q=aT_{3L}+\frac{b}{\sqrt{3}}T_{8L}+ \frac{c}{\sqrt{6}}T_{15L}+ XI_4, 
\end{equation} 
where $T_{iL}=\lambda_{iL}/2$ ($\lambda_{iL}$ are the Gell-Mann matrices for $SU(4)_L$ normalized as  Tr$(\lambda_i\lambda_j)=2\delta_{ij}$), 
$I_4=Dg(1,1,1,1)$ is the diagonal $4\times 4$ unit matrix, and $a=1$ gives the usual isospin of the electroweak interaction. These possibilities are: $b= c = 1$ and $b = 1, c = -2$, which become a convenient classification scheme for these type of models. Four of the identified models without exotic electric charges are three-family models. Two of them are models for which $b= c = 1$ and have been analyzed in Refs.~\cite{pgs} and \cite{swz}. The other two models belong to the class for which $b = 1, c = -2$ and have been studied in Refs.~\cite{spp} and \cite{sen}.

3-4-1 models containing exotic electric charges have been also considered in the literature \cite{su41,su42}. In this case a particular embedding of the SM gauge group into $SU(3)_c\otimes SU(4)_L\otimes U(1)_X$ depends on the physical motivation of the model to be constructed. The model in Ref.~\cite{su41}, for example, has been proposed with the goal of including right-handed neutrinos in the fermion spectrum from the start.

Models based on the $SU(3)_c\otimes SU(4)_L\otimes U(1)_X$ gauge symmetry predict the existence of three massive neutral currents which mix with each other. They are, the usual neutral current of the SM associated to the $Z$ gauge boson, and two new associated to the gauge bosons $Z^\prime$ and $Z^{\prime \prime}$. Unlike models containing exotic electric charges, for models with only ordinary electric charges the mixing can be constrained to occur between $Z$ and $Z^\prime$ only \cite{pgs,swz,spp}. This fact produces an enormous simplification in the study of the low energy deviations of the $Z$ couplings to the SM families \cite{little1}. On the other hand, after the breakdown of the 3-4-1 symmetry down to $SU(3)_c \otimes U(1)_Q$, the left-handed couplings of quarks to the SM $Z$ boson remain flavor conserving (at low energies the model coincides with the SM); but, since anomaly cancellation among generations forces one family of quarks to transform differently from the other two, the left-handed couplings of quarks both to $Z^\prime$ and $Z^{\prime \prime}$ are, in general, not flavor diagonal. Consequently, flavor changing neutral currents (FCNC) arise.

With a family of quarks transforming differently under $SU(4)_L\otimes U(1)_X$, we have three possible assignments of weak eigenstates into mass eigenstates. As we will see, the phenomenological implications of the model will depend on the choice of the quark family being different.

In this work we will constrain ourselves to the case where the mixing occurs between $Z$ and $Z^\prime$ only, with $Z^{\prime \prime}\equiv Z_3$ being a mass eigenstate \cite{little1}. Then, for two different 3-4-1 models, one of them representative of the $b=c=1$ class and the other one representative of the $b=1,\; c=-2$ class, we do a $\chi^2$ fit to Z-pole observables and atomic parity violation (APV) data in order to constraint the mixing angle $\theta$ between $Z$ and $Z^\prime$ and the mass scale $M_{Z_2}$ of the corresponding physical new neutral gauge boson. The main purpose will be to examine how much these bounds depend on the three different assignments of quark gauge eigenstates into mass eigenstates. Next, for the same two different 3-4-1 models and for the three possible assignments, we will impose constraints on the parameters $M_{Z_2}$ and $M_{Z_3}$ coming from neutral meson mixing in the analysis of the FCNC effects present in the models. The outcome of the analysis will be then used to establish which quark family must transform differently in order to keep the lower bounds on $M_{Z_2}$ and $M_{Z_3}$ as low as possible. 

This paper is organized as follows. In the next section we review the 3-4-1 models to be studied with the emphasis done on the aspects relevant to the analysis proposed in the previous paragraph. In Sec.~\ref{sec:sec3} we use electroweak measurements at the Z-pole, APV data and experimental input from neutral meson mixing in order to obtain family-dependent bounds on the parameters of the models. Finally, in the last section, we summarize and present our conclusions.

\section{\label{sec:sec2}3-4-1 models}
As stated, we will consider models without exotic electric charges based on the 3-4-1 symmetry, each one characterized by the values of the parameters $b$ and $c$ in the electric charge operator in Eq.~(\ref{ch}), namely: $b=c=1$ or $b=1,\; c=-2$. There exists two anomaly-free models of each type \cite{sap}. We will select one representative model of the $b=c=1$ class and one representative model of the $b=1,\; c=-2$ class.

In what follows we assume the symmetry breaking pattern
\begin{eqnarray}\nonumber
SU(3)_c\otimes SU(4)_L\otimes & U(1)_X & \\ \nonumber
& \stackrel{V^\prime}{\longrightarrow}
& SU(3)_c\otimes SU(3)_L\otimes U(1)_Z \\ \nonumber 
& \stackrel{V}{\longrightarrow} & SU(3)_c\otimes SU(2)_L\otimes U(1)_Y \\ \label{break}
& \stackrel{v+v^\prime}{\longrightarrow} & SU(3)_c\otimes U(1)_Q,
\end{eqnarray}
where $SU(3)_c\otimes SU(3)_L\otimes U(1)_Z$ refers to the three-family 3-3-1 structure introduced in Ref.~\cite{331}, and $V$, $V^\prime$, $v$, $v^\prime$ are the vacuum expectation values of four Higgs 4-plets which will be specified for each model below. We impose the hierarchy $V\sim V^\prime >> v\sim v^\prime \simeq 174$~GeV.

The gauge couplings $g_4$ and $g_X$, associated with the groups $SU(4)_L$ and $U(1)_X$, respectively, are defined through the covariant derivative for 4-plets as: $iD^\mu = i\partial^\mu - g_4 \lambda_{L\alpha}A^\mu_\alpha/2 - g_X X B^\mu$.

\subsection{\label{sec:sub2a}Model A: $b=c=1$}
This is Model B in Ref.~\cite{sap}. It has the anomaly-free fermion content displayed in Table~\ref{tab:1} where $i=1,2$ and $\alpha=1,2,3$ are family indexes and the numbers in parentheses refer to the $[SU(3)_C, SU(4)_L, U(1)_X]$ quantum numbers, respectively. $U_i$ and $U^\prime_i$ are exotic quarks of electric charge $2/3$, while $D_3$ and $D^\prime_3$ are exotic quarks of electric charge $-1/3$. $E^-_\alpha$ and $E^{\prime -}_\alpha$ are exotic electrons. Notice that universality for the known leptons in the three families is present at the tree level in the weak basis. So, FCNC do not occur in the lepton sector, up to possible mixing with the exotic fields.

From Table~\ref{tab:1} we can identify three different realizations in the mass basis as shown in Table~\ref{tab:2}.

The symmetry breaking and masses for all the fermion fields (except for $\nu^0_{e \alpha}$) are produced by the set of Higgs scalars \cite{swz}
\begin{eqnarray}\nonumber 
\langle\phi^T_1\rangle&=&\langle(\phi^0_1,\phi^+_1,\phi^{\prime +}_1,\phi^{\prime\prime +}_1)\rangle=(v,0,0,0)\sim[1,4^*,3/4], \\ \nonumber
\langle\phi^T_2\rangle&=&\langle(\phi^-_2,\phi^0_2,\phi^{\prime 0}_2,\phi^{\prime\prime 0}_2)\rangle=(0,v',0,0)\sim[1,4^*,-1/4], \\ \nonumber
\langle\phi^T_3\rangle&=&\langle(\phi^-_3,\phi^0_3,\phi^{\prime 0}_3,\phi^{\prime\prime 0}_3)\rangle=(0,0,V,0)\sim[1,4^*,-1/4], \\ \label{scalara}
\langle\phi^T_4\rangle&=&\langle(\phi^-_4,\phi^0_4,\phi^{\prime 0}_4,\phi^{\prime\prime 0}_4)\rangle=(0,0,0,V')\sim[1,4^*,-1/4].
\end{eqnarray}

\begin{table}
\caption{\label{tab:1}Anomaly-free fermion content of Model A.}
\begin{ruledtabular}
\begin{tabular}{ccccc}
$Q_{iL}=\left(\begin{array}{c}d_i\\u_i\\U_i\\U^{\prime}_i 
\end{array}\right)_L$ & 
$d^c_{iL}$ & $u^c_{iL}$ & $U^c_{iL}$ & $U^{\prime c}_{iL}$ \\
$[3,4^*,\frac{5}{12}]$ & $[3^*,1,{1\over 3}]$ & $[3^*,1,-{2\over 3}]$
& $[3^*,1,-{2\over 3}]$ & $[3^*,1,-{2\over 3}]$ \\ \hline
$Q_{3L}=\left(\begin{array}{c}u_3\\d_3\\D_3\\D^{\prime}_3 \end{array}\right)_L$ &
$u^c_{3L}$ & $d^c_{3L}$ & $D^c_{3L}$ & $D^{\prime c}_{3L}$ \\
$[3,4,-\frac{1}{12}]$ & $[3^*,1,-{2\over 3}]$ & $[3^*,1,{1\over 3}]$ & $[3^*,1,{1\over 3}]$ & $[3^*,1,{1\over 3}]$ \\ \hline 
$L_{\alpha L}=\left(\begin{array}{c} \nu^0_{e \alpha}\\ e^-_\alpha\\  
E^{-}_\alpha \\ E^{\prime -}_\alpha \end{array}\right)_L$ & 
$e^+_{\alpha L}$ & $E^+_{\alpha L}$ & $E^{\prime +}_{\alpha L}$ & $$\\
$[1,4,-{3\over 4}]$ & $[1,1,1]$ & $[1,1,1]$ & $[1,1,1]$ & $$ \\
\end{tabular}
\end{ruledtabular}
\end{table}

\begin{table}
\caption{\label{tab:2}Three different assignments of symmetry eigenstates into mass eigenstates for Model A.}
\begin{ruledtabular}
\begin{tabular}{ccc}
$\mbox{Assignment A1}$ & $\mbox{Assignment A2}$ & $\mbox{Assignment A3}$ \\ \hline
$Q_{iL}=\left(\begin{array}{c}d,s\\u,c\\U_1,U_2\\U^{\prime}_1,U^{\prime}_2
\end{array}\right)_L$ & 
$Q_{iL}=\left(\begin{array}{c}d,b\\u,t\\U_1,U_3\\U^{\prime}_1,U^{\prime}_3 \end{array}\right)_L$ & 
$Q_{iL}=\left(\begin{array}{c}s,b\\c,t\\U_2,U_3\\U^{\prime}_2,U^{\prime}_3 \end{array}\right)_L$  \\
$Q_{3L}=\left(\begin{array}{c}t\\b\\D_3\\D^{\prime}_3 \end{array}\right)_L$ &
$Q_{3L}=\left(\begin{array}{c}c\\s\\D_2\\D^{\prime}_2 \end{array}\right)_L$ & $Q_{3L}=\left(\begin{array}{c}u\\d\\D_1\\D^{\prime}_1 \end{array}\right)_L$ \\
\end{tabular}
\end{ruledtabular}
\end{table}

When the 3-4-1 symmetry is broken to the SM, we get the gauge matching conditions
\begin{eqnarray}
g_4 = g, \qquad&\mbox{and}&\qquad\frac{1}{g^{\prime 2}}=\frac{1}{g^2_X}+\frac{1}{2g^2}, \label{matchA}
\end{eqnarray}
where $g$ and $g^\prime$ are the gauge coupling constants of the $SU(2)_L$ and $U(1)_Y$ groups of the SM, respectively.

Mixing between ordinary and exotic fermions and violation of unitarity of the Cabibbo-Kobayashi-Maskawa (CKM) mixing matrix can be avoided by introducing an anomaly-free discrete $Z_2$ symmetry with assignments of $Z_2$ charge $q_Z$ given by
\begin{eqnarray} \nonumber
q_Z(Q_{\alpha L}, u^c_{\alpha L}, d^c_{\alpha L}, L_{\alpha L}, e^c_{\alpha L}, \phi_1, \phi_2)&=& 0, \\ \label{z2a}
q_Z(U^c_{iL}, U^{\prime c}_{iL}, D^c_{3L}, D^{\prime c}_{3L}, E^c_{\alpha L}, E^{\prime c}_{\alpha L}, \phi_3, \phi_4)&=& 1,
\end{eqnarray}

After the symmetry breaking, the Yukawa couplings allowed by the gauge invariance and the $Z_2$ symmetry produce for up- and down-type quarks, in the basis $(u_1,u_2,u_3,$ $U_1,U_2,U'_1,U'_2)$ and $(d_1,d_2,d_3,D_3,D'_3)$, respectively, block diagonal mass matrices of the form
\begin{equation}\label{mupa}
M_{uU}=\left(\begin{array}{cc}
M^u_{3\times 3} &  0\\
0 &  M^U_{4\times 4}\end{array}\right), 
\end{equation}
and
\begin{equation}\label{mdowna}
M_{dD}=\left(\begin{array}{cc}
M^d_{3\times 3} &  0\\
0 &  M^D_{2\times 2}\end{array}\right),
\end{equation}
and similarly for the charged leptons.
 
For our purposes, we will be mainly interested in the neutral gauge boson sector which consists of four physical fields: the massless photon $A_\mu$ and the massive gauge bosons $Z_\mu$, $Z^\prime_\mu$ and $Z^{\prime \prime}_\mu$. In terms of the electroweak basis, they are given by
\begin{eqnarray} \nonumber
A^\mu&=&S_W A_3^\mu + C_W Y^\mu\; , \\ \nonumber  
Z^\mu&=& C_W A_3^\mu - S_W Y^\mu \; , \\ \nonumber 
Z'^\mu&=&\sqrt{\frac{2}{3}}(1-T_W^2/2)^{1/2}\left(A_8^\mu+
\frac{A_{15}^\mu}{\sqrt{2}}\right)-\frac{T_W}{\sqrt{2}}B^\mu, \\ \label{fzzpa}
Z''^\mu&=& A_8^\mu / \sqrt{3}-\sqrt{2/3}A_{15}^\mu, 
\end{eqnarray}
where
\begin{equation}\label{ya}
Y^\mu=\frac{T_W}{\sqrt{3}}\left(A_8^\mu+
\frac{A_{15}^\mu}{\sqrt{2}}\right)+(1-T_W^2/2)^{1/2}B^\mu
\end{equation}
is the field to be identified as the $Y$ hypercharge associated with the SM abelian gauge boson. $T_W=S_W/C_W$, where $S_W$ is the sine of the electroweak mixing angle defined as $S_W=\sqrt{2}g_X /\sqrt{3 g^2_X + 2 g^2_4}$, and $C_W=\sqrt{1-S^2_W}$.

Since we are interested in the low energy phenomenology of the model, we can choose $V\simeq V^\prime$. Moreover we will consider the particular case $v\simeq v^\prime$, for which the current $Z^{\prime \prime \mu}\equiv Z^\mu_3$ decouples from the other two and acquires a squared mass $M^2_{Z_3}=(g^2_4/2)V^2$ \cite{little1}. The remaining mixing between $Z_\mu$ and $Z^\prime_\mu$ is parametrized by the mixing angle $\theta$ as
\begin{eqnarray}\nonumber
Z_1^\mu&=&Z^\mu \cos\theta+Z^{\prime\mu} \sin\theta \; ,\\ \label{mixa}
Z_2^\mu&=&-Z^\mu \sin\theta+Z^{\prime\mu} \cos\theta, 
\end{eqnarray} 
where $Z_1^\mu$ and $Z_2^\mu$ are the mass eigenstates and
\begin{equation} \label{tana} \tan(2\theta) = \frac{2 \sqrt{2} \delta v^2 S^3_W}
{2 \delta^2[v^2(S^4_W+C^4_W)+V^2 C^4_W]-v^2 S^2_W}, 
\end{equation}
with $\delta=g_X/(2g_4)$.

The Lagrangian for neutral currents can be written as 
$-{\cal L}^{\mathrm{NC}} = eA^\mu J_\mu(\mathrm{EM})+(g_4/{C_W})Z^\mu J_\mu(Z)
+ (g_X/\sqrt{2})Z'^\mu$ $J_\mu(Z') + (g_4/2)Z''^\mu J_\mu(Z'')$,
with
\begin{eqnarray}\nonumber
J_\mu(\mathrm{EM})&=&{2\over 3}[\bar{u}_3\gamma_\mu u_3 +\sum_{i=1}^2(\bar{u}_i\gamma_\mu u_i + \bar{U}_i\gamma_\mu U_i \\ \nonumber
& & + \bar{U'}_i\gamma_\mu U'_i)] \\ \nonumber
& & -{1\over3}(\bar{d}_3\gamma_\mu d_3 + \bar{D}_3\gamma_\mu D_3 + \bar{D'}_3\gamma_\mu D'_3 \\ \nonumber
& & +\sum_{i=1}^2\bar{d}_i\gamma_\mu d_i) \\ \nonumber
& & -\sum_{\alpha=1}^3(\bar{e}^-_\alpha\gamma_\mu e^-_\alpha + \bar{E}^-_\alpha\gamma_\mu E^-_\alpha + \bar{E'}^-_\alpha\gamma_\mu E^{\prime -}_\alpha) \\ \label{emca}
&=&\sum_f q_f\bar{f}\gamma_\mu f,
\end{eqnarray}
\begin{equation}\label{zca}
J_\mu(Z)=J_{\mu,L}(Z)-S^2_WJ_\mu(\mathrm{EM}),
\end{equation}

\begin{equation}\label{zpca} 
J_\mu(Z')=J_{\mu,L}(Z')-T_WJ_\mu(\mathrm{EM}),
\end{equation}
\begin{eqnarray}\nonumber
J_\mu(Z'')&=&-\bar{D}_{3L}\gamma_\mu D_{3L} + \bar{D'}_{3L}\gamma_\mu D'_{3L} \\ \nonumber
& & +\sum_{i=1}^2(\bar{U}_{iL}\gamma_\mu 
U_{iL}-\bar{U'}_{iL}\gamma_\mu U'_{iL})\\ \label{zppca}
& & -\sum_{\alpha=1}^3(\bar{E^-}_{\alpha L}\gamma_\mu E^-_{\alpha L}
-\bar{E}^{\prime -}_{\alpha L}\gamma_\mu E^{\prime -}_{\alpha L}),
\end{eqnarray}
\noindent 
where $e=gS_W=g_4S_W=g_XC_W\sqrt{1-T_W ^2/2}>0$ is the electric charge, $q_f$ is the electric charge of the fermion $f$ in units of $e$ and 
$J_\mu(EM)$ is the electromagnetic current. It is important to remark that in this model the $Z''_\mu$ current couples only to exotic fields as can be seen from Eq.~(\ref{zppca}). 

The left-handed currents in $J_\mu(Z)$ and $J_\mu(Z')$ are
\begin{eqnarray} \nonumber
J_{\mu,L}(Z)&=&{1\over 2} [\bar{u}_{3L}\gamma_\mu u_{3L}-\bar{d}_{3L}\gamma_\mu d_{3L} \\ \nonumber
& & -\sum_{i=1}^2(\bar{d}_{iL}\gamma_\mu d_{iL}-\bar{u}_{iL}\gamma_\mu u_{iL})\\ \nonumber
& & +\sum_{\alpha=1}^3(\bar{\nu}_{e\alpha L}\gamma_\mu \nu_{e\alpha L} - \bar{e}^-_{\alpha L}\gamma_\mu e^-_{\alpha L})] \\ \label{lhzca} 
&=&\sum_f T_{4f}\bar{f}_L\gamma_\mu f_L ,
\end{eqnarray}

\begin{table*}
\caption{\label{tab:3}The $Z_1^\mu\longrightarrow \bar{f}f$ couplings for Model A.}
\begin{ruledtabular}
\begin{tabular}{lcc}
$f$ & $g^{(A)}(f)_{1V}$ & $g^{(A)}(f)_{1A}$ \\ \hline
$u_3$ & $({1\over 2}-{4S_W^2 \over 3})(\cos\theta +
\Upsilon \sin\theta) $
& ${1\over 2} (\cos\theta + \Upsilon \sin\theta) $ \\ 
$d_3$ & $(-{1\over
2}+{2S_W^2\over 3})\cos\theta+{1\over 2}\Upsilon (C^2_W+\frac{S^2_W}{3})\sin\theta $ 
& $-{1\over 2}(\cos\theta - \Upsilon C_{2W}\sin\theta )$ \\
$D_3$ & ${2S_W^2\over 3}\cos\theta-{1\over 2}\Upsilon (1 
- {7 S_W^2 \over 3})\sin\theta $ &
$-{1\over 2}\Upsilon C_W^2\sin\theta $ \\  
$D'_3$ & ${2S_W^2\over 3}\cos\theta-{1\over 2}\Upsilon (1 
- {7 S_W^2 \over 3})\sin\theta $ &
$-{1\over 2}\Upsilon C_W^2\sin\theta $ \\
$d_{1,2}$ & $(-{1\over
2}+{2S_W^2\over 3})(\cos\theta +\Upsilon \sin\theta) $ & $-\frac{1}{2}(\cos\theta 
+ \Upsilon\sin\theta) $ \\
$u_{1,2}$& $({1\over 2}-{4S_W^2 \over 3})\cos\theta -
{1\over 2}\Upsilon (C^2_W +{5 S_W^2\over 3})\sin\theta $
& ${1\over 2} (\cos\theta - \Upsilon C_{2W}\sin\theta) $ \\ 
$U_{1,2}$& $-{4S_W^2\over 3}\cos\theta
+\frac{1}{2}\Upsilon (1-{11\over 3}S_W^2)\sin\theta $ &
$\frac{1}{2}\Upsilon C_W^2\sin\theta $ \\
$U'_{1,2}$& $-{4S_W^2\over 3}\cos\theta
+\frac{1}{2}\Upsilon (1-{11\over 3}S_W^2)\sin\theta $ &
$\frac{1}{2}\Upsilon C_W^2\sin\theta $ \\  
$\nu_{1,2,3}$& ${1\over 2}(\cos\theta
+\Upsilon\sin\theta)$ & $ {1\over 2}(\cos\theta
+\Upsilon\sin\theta)$\\
$e^-_{1,2,3}$ &
$(-{1\over 2}+2 S^2_W)\cos\theta+\Upsilon ({1\over 2}+ S^2_W)\sin\theta $ & $ -{1\over 2}(\cos\theta - \Upsilon C_{2W}\sin\theta )$ \\ 
$E^-_{1,2,3}$ & $2 S_W^2\cos\theta + \frac{1}{2}\Upsilon (-1+5 S^2_W)\sin\theta $ &
$-\frac{1}{2}\Upsilon C^2_W \sin\theta $ \\ 
$E^{\prime -}_{1,2,3}$ &
$2 S_W^2\cos\theta + \frac{1}{2}\Upsilon (-1+5 S^2_W)\sin\theta $ &
$-\frac{1}{2}\Upsilon C^2_W\sin\theta $ \\
\end{tabular}
\end{ruledtabular}
\end{table*}

\begin{table*}
\caption{\label{tab:4}The $Z_2^\mu\longrightarrow \bar{f}f$ couplings for Model A.}
\begin{ruledtabular}
\begin{tabular}{lcc}
$f$ & $g^{(A)}(f)_{2V}$ & $g^{(A)}(f)_{2A}$ \\ \hline
$u_3$ & $-({1\over 2}-{4S_W^2 \over 3})(\sin\theta -
\Upsilon \cos\theta) $
& ${1\over 2} (-\sin\theta + \Upsilon \cos\theta) $ \\ 
$d_3$ & $({1\over 2}-{2S_W^2\over 3})\sin\theta+{1\over 2}\Upsilon (C^2_W+\frac{S^2_W}{3})\cos\theta $ 
& ${1\over 2}(\sin\theta + \Upsilon C_{2W}\cos\theta )$ \\
$D_3$ & $-{2S_W^2\over 3}\sin\theta-{1\over 2}\Upsilon (1 
- {7 S_W^2 \over 3})\cos\theta $ &
$-{1\over 2}\Upsilon C_W^2\cos\theta $ \\  
$D'_3$ & $-{2S_W^2\over 3}\sin\theta-{1\over 2}\Upsilon (1 
- {7 S_W^2 \over 3})\cos\theta $ &
$-{1\over 2}\Upsilon C_W^2\cos\theta $ \\
$d_{1,2}$ & $({1\over
2}-{2S_W^2\over 3})(\sin\theta -\Upsilon \cos\theta) $ & 
$-\frac{1}{2}(-\sin\theta 
+ \Upsilon\cos\theta) $ \\
$u_{1,2}$& $-({1\over 2}-{4S_W^2 \over 3})\sin\theta -
{1\over 2}\Upsilon (C^2_W +{5 S_W^2\over 3})\cos\theta $
& $-{1\over 2} (\sin\theta + \Upsilon C_{2W}\cos\theta) $ \\ 
$U_{1,2}$& ${4S_W^2\over 3}\sin\theta
+\frac{1}{2}\Upsilon (1-{11\over 3}S_W^2)\cos\theta $ &
$\frac{1}{2}\Upsilon C_W^2\cos\theta $ \\
$U'_{1,2}$& ${4S_W^2\over 3}\sin\theta
+\frac{1}{2}\Upsilon (1-{11\over 3}S_W^2)\cos\theta $ &
$\frac{1}{2}\Upsilon C_W^2\cos\theta $ \\  
$\nu_{1,2,3}$& ${1\over 2}(-\sin\theta
+\Upsilon\cos\theta)$ & $ {1\over 2}(-\sin\theta
+\Upsilon\cos\theta)$\\
$e^-_{1,2,3}$ &
$({1\over 2}-2 S^2_W)\sin\theta+\Upsilon ({1\over 2}+ S^2_W)\cos\theta $ & $ {1\over 2}(\sin\theta + \Upsilon C_{2W}\cos\theta )$ \\ 
$E^-_{1,2,3}$ & $-2 S_W^2\sin\theta + \frac{1}{2}\Upsilon (-1+5 S^2_W)\cos\theta $ &
$-\frac{1}{2}\Upsilon C^2_W \cos\theta $ \\ 
$E^{\prime -}_{1,2,3}$ &
$-2 S_W^2\sin\theta + \frac{1}{2}\Upsilon (-1+5 S^2_W)\cos\theta $ &
$-\frac{1}{2}\Upsilon C^2_W\cos\theta $ \\
\end{tabular}
\end{ruledtabular}
\end{table*}

\begin{eqnarray}\nonumber
J_{\mu,L}(Z')&=&(2T_{W})^{-1}\lbrace(1+T^2_W)\bar{u}_{3L}\gamma_\mu u_{3L} \\ \nonumber 
& & +(1-T^2_W)\bar{d}_{3L}\gamma_\mu d_{3L} \\ \nonumber
& & -\bar{D}_{3L}\gamma_\mu D_{3L}-\bar{D'}_{3L}\gamma_\mu D'_{3L} \\ \nonumber
& & -\sum_{i=1}^2\lbrack (1+T^2_W)\bar{d}_{iL}\gamma_\mu d_{iL} \\ \nonumber
& & +(1-T^2_W)\bar{u}_{iL}\gamma_\mu u_{iL} \\ \nonumber
& & -\bar{U}_{iL}\gamma_\mu U_{iL}-\bar{U'}_{iL}\gamma_\mu U'_{iL}\rbrack \\ \nonumber 
& & +\sum_{\alpha=1}^3\lbrack (1+T^2_W)\bar{\nu}_{\alpha L}\gamma_\mu \nu_{\alpha L} \\ \nonumber
& & +(1-T^2_W)\bar{e}^-_{\alpha L}\gamma_\mu e^-_{\alpha L} \\ \nonumber
& & -\bar{E}^-_{\alpha L}\gamma_\mu E^-_{\alpha L} -\bar{E}^{\prime -}_{\alpha L}\gamma_\mu E^{\prime -}_{\alpha L}\rbrack \rbrace
\\ \label{lhzpca}
&=&\sum_f T'_{4f}\bar{f}_L\gamma_\mu f_L,
\end{eqnarray}
where $T_{4f}=Dg(1/2,-1/2,0,0)$ is the third component of the weak isospin
and $T'_{4f}=(1/2T_W)Dg(1+T^2_W, 1-T^2_W, -1, -1)$= $T_W\lambda_3/2 +(1/T_W)(\lambda_8/\sqrt{3}+\lambda_{15}/\sqrt{6})$ is a convenient $4\times 4$ diagonal matrix, both of them acting on the representation 4 of $SU(4)_L$. The current $J_\mu(Z)$ is clearly recognized as the generalization of the neutral current of the SM. Thus, we identify $Z_\mu$ as the neutral gauge boson of the SM.

From Eq.~(\ref{lhzpca}) we can see that the left-handed couplings of $Z^\prime$ to one family of quarks are different from the couplings to the other two. This induces FCNC at the tree level transmitted by the $Z^\prime$ boson.

The couplings between the mass eigenstates $Z_1^\mu$, $Z_2^\mu$ and the fermion fields are obtained for this model from the Hamiltonian
\begin{equation} \nonumber
{\cal H}^{\mathrm{NC}}=\frac{g_4}{2C_W}\sum_{i=1}^2Z_i^\mu\sum_f\{\bar{f}\gamma_\mu [g^{(A)}(f)_{iV}-g^{(A)}(f)_{iA}\gamma_5]f\},
\end{equation}

The expressions for $g^{(A)}_{iV},\; g^{(A)}_{iA}$ ($i=1,2$) are listed in Tables~\ref{tab:3} and \ref{tab:4}, where $\Upsilon=1/\sqrt{2-3S^2_W}$ and $C_{2W} = C^2_W - S^2_W$.

\subsection{\label{sec:sub2b}Model B: $b=1,\; c=-2$}

This is Model E in Ref.~\cite{sap}. For this model the anomaly-free fermion structure is given in Table~\ref{tab:5}, which shows that also in this case universality for the ordinary leptons in the three families is present in the weak basis. From this Table we can recognize the three different realizations in the mass basis shown in Table~\ref{tab:6}.

The symmetry breaking and masses for all the fermion fields (except for the neutral leptons) are achieved with the set of Higgs scalars \cite{spp}
\begin{eqnarray}\nonumber 
\langle\phi^T_1\rangle&=&\langle(\phi^-_1,\phi^0_1,\phi^{\prime 0}_1,\phi^{\prime -}_1)\rangle=(0,v,0,0)\sim[1,4^*,-1/2], \\ \nonumber
\langle\phi^T_2\rangle&=&\langle(\phi^-_2,\phi^0_2,\phi^{\prime 0}_2,\phi^{\prime -}_2)\rangle=(0,0,V,0)\sim[1,4^*,-1/2], \\ \nonumber
\langle\phi^T_3\rangle&=&\langle(\phi^0_3,\phi^-_3,\phi^{\prime -}_3,\phi^{\prime 0}_3)\rangle=(v',0,0,0)\sim[1,4,-1/2], \\ \label{scalarb}
\langle\phi^T_4\rangle&=&\langle(\phi^0_4,\phi^-_4,\phi^{\prime -}_4,\phi^{\prime 0}_4)\rangle=(0,0,0,V')\sim[1,4,-1/2].
\end{eqnarray}

The gauge matching conditions now read
\begin{eqnarray}
g_4 = g, \qquad&\mbox{and}&\qquad\frac{1}{g^{\prime 2}}=\frac{1}{g^2_X}+\frac{1}{g^2}. \label{matchB}
\end{eqnarray}

\begin{table}
\caption{\label{tab:5}Anomaly-free fermion content of Model B.}
\begin{ruledtabular}
\begin{tabular}{ccccc}
$Q_{iL}=\left(\begin{array}{c}u_i\\d_i\\D_i\\U_i \end{array}\right)_L$ & 
$u^c_{iL}$ & $d^c_{iL}$ & $D^c_{iL}$ & $U^{c}_{iL}$ \\
$[3,4,\frac{1}{6}]$ & $[3^*,1,-{2\over 3}]$ & $[3^*,1,{1\over 3}]$
& $[3^*,1,{1\over 3}]$ & $[3^*,1,-{2\over 3}]$ \\ \hline
$Q_{3L}=\left(\begin{array}{c}d_3\\u_3\\U_3\\D_3 
\end{array}\right)_L$ &
$d^c_{3L}$ & $u^c_{3L}$ & $U^c_{3L}$ & $D^{\prime c}_{3L}$ \\
$[3,4^*,\frac{1}{6}]$ & $[3^*,1,-{2\over 3}]$ & $[3^*,1,{1\over 3}]$ & $[3^*,1,{1\over 3}]$ & $[3^*,1,{1\over 3}]$ \\ \hline 
$L_{\alpha L}=\left(\begin{array}{c} e^-_\alpha\\ \nu^0_{e \alpha}\\ 
N^0_\alpha \\ E^{-}_\alpha \end{array}\right)_L $ & 
$e^+_{\alpha L}$ & $E^+_{\alpha L}$ & $$ & $$ \\
$[1,4^*,-{1\over 2}]$ & $[1,1,1]$ & $[1,1,1]$ & $$ & $$ \\
\end{tabular}
\end{ruledtabular}
\end{table}

\begin{table}
\caption{\label{tab:6}Three different assignments of symmetry eigenstates into mass eigenstates for Model B.}
\begin{ruledtabular}
\begin{tabular}{ccc}
$\mbox{Assignment B1}$ & $\mbox{Assignment B2}$ & $\mbox{Assignment B3}$ \\ \hline
$Q_{iL}=\left(\begin{array}{c}u,c\\d,s\\D_1,D_2\\U_1,U_2 \end{array}\right)_L$ & 
$Q_{iL}=\left(\begin{array}{c}u,t\\d,b\\D_1,D_3\\U_1,U_3 \end{array}\right)_L$ & $Q_{iL}=\left(\begin{array}{c}c,t\\s,b\\D_2,D_3\\U_2,U_3 \end{array}\right)_L$  \\
$Q_{3L}=\left(\begin{array}{c}b\\t\\U_3\\D_3 
\end{array}\right)_L$ &
$Q_{3L}=\left(\begin{array}{c}s\\c\\U_2\\D_2 
\end{array}\right)_L$ & $Q_{3L}=\left(\begin{array}{c}d\\u\\U_1\\D_1 
\end{array}\right)_L$ \\
\end{tabular}
\end{ruledtabular}
\end{table}

For this model the discrete $Z_2$ symmetry assigns charges $q_Z$ according to
\begin{eqnarray} \nonumber
q_Z(Q_{\alpha L}, u^c_{\alpha L}, d^c_{\alpha L}, L_{\alpha L}, e^c_{\alpha L}, \phi_1, \phi_3)&=& 0, \\ \label{z2b}
q_Z(U^c_{\alpha L}, D^c_{\alpha L}, E^c_{\alpha L}, \phi_2, \phi_4)&=& 1,
\end{eqnarray}
and the resulting block diagonal form of the mass matrices for up- and down-type quarks, in the basis $(u_1,u_2,u_3,U_1,$ $U_2,U_3)$ and $(d_1,d_2,d_3,D_1,D_2,D_3)$ respectively, are
\begin{equation}\label{mupb}
M_{uU}=\left(\begin{array}{cc}
M^u_{3\times 3} &  0\\ 
0 &  M^U_{3\times 3}\end{array}\right),
\end{equation} 
and
\begin{equation}\label{mdownb}
M_{dD}=\left(\begin{array}{cc}
M^d_{3\times 3} &  0\\
0 &  M^D_{3\times 3}\end{array}\right),
\end{equation}
and similarly for the charged leptons.
 
The photon field $A_\mu$ and the massive neutral gauge bosons are
\begin{eqnarray} \nonumber
A^\mu&=&S_W A_3^\mu \nonumber \\
& & + C_W\left[\frac{T_W}{\sqrt{3}}\left(A_8^\mu-
2\frac{A_{15}^\mu}{\sqrt{2}}\right)+(1-T_W^2)^{1/2}B^\mu\right]\; , \nonumber \\  
Z^\mu&=& C_W A_3^\mu \nonumber \\
& & - S_W\left[\frac{T_W}{\sqrt{3}}\left(A_8^\mu-
2\frac{A_{15}^\mu}{\sqrt{2}}\right)+(1-T_W^2)^{1/2}B^\mu\right] \; , \nonumber \\ \nonumber
Z'^\mu&=&\frac{1}{\sqrt{3}}(1-T_W^2)^{1/2}\left(A_8^\mu-
2\frac{A_{15}^\mu}{\sqrt{2}}\right)-T_W B^\mu, \\ \label{fzzpb}
Z''^\mu&=& 2A_8^\mu /\sqrt{6}+A_{15}^\mu/\sqrt{3},
\end{eqnarray}
\noindent
from which we identify the $Y$ hypercharge associated with the SM $U(1)_Y$ gauge boson as
\begin{equation}\label{yb}
Y^\mu=\frac{T_W}{\sqrt{3}}\left(A_8^\mu-
2\frac{A_{15}^\mu}{\sqrt{2}}\right)+(1-T_W^2)^{1/2}B^\mu.
\end{equation}

Also in this model, for $V\simeq V^\prime$ and $v\simeq v^\prime$, the field $Z^{\prime \prime \mu}\equiv Z^\mu_3$ decouples and acquires a mass $M^2_{Z_3}=(g^2_4/2)(V^2$ $+v^2)$. The mixing angle between $Z_\mu$ and $Z^{\prime}_\mu$ is given by
\begin{equation} \label{tanb} 
\tan(2\theta) =  \frac{v^2 S_W^2 \sqrt{C_{2W}}}
{v^2(1+S_W^2)^2 + V^2 C_W^4 - 2v^2}.
\end{equation}
\noindent 
where the sine of the electroweak mixing angle is defined as $S_W=g_X/\sqrt{2g^2_X+g^2_4}$.

The neutral currents are
\begin{eqnarray}\nonumber
J_\mu(\mathrm{EM})&=&{2\over 3}\lbrack \sum_{i=1}^2(\bar{u}_i\gamma_\mu u_i+
\bar{U}_i\gamma_\mu U_i)
+\bar{u}_3\gamma_\mu u_3 \\ \nonumber
& & +\bar{U}_3\gamma_\mu U_3 \rbrack -{1\over3}\lbrack \sum_{i=1}^2(\bar{d}_i\gamma_\mu d_i+ \bar{D}_i\gamma_\mu D_i) \\ \nonumber & & +\bar{d}_3\gamma_\mu d_3+\bar{D}_3\gamma_\mu D_3 \rbrack \\ \nonumber
& & -\sum_{\alpha=1}^3\bar{e}^-_\alpha\gamma_\mu e^-_\alpha -\sum_{\alpha=1}^3\bar{E}^-_\alpha\gamma_\mu E^-_\alpha \\ \label{emcb}
&=&\sum_f q_f\bar{f}\gamma_\mu f, 
\end{eqnarray}

\begin{equation}\label{zcb}
J_\mu(Z)=J_{\mu,L}(Z)-S^2_WJ_\mu(\mathrm{EM}),
\end{equation}
\begin{equation}\label{zpcb}
J_\mu(Z')=J_{\mu,L}(Z')-T_WJ_\mu(\mathrm{EM}),
\end{equation}

\begin{eqnarray}\nonumber
J_\mu(Z'')&=&\sum_{i=1}^2(\bar{u}_{iL}\gamma_\mu u_{iL}+\bar{d}_{iL}\gamma_\mu d_{iL}-\bar{D}_{iL}\gamma_\mu 
D_{iL} \\ \nonumber
& & -\bar{U}_{iL}\gamma_\mu U_{iL})-\bar{d}_{3L}\gamma_\mu d_{3L}-\bar{u}_{3L}\gamma_\mu u_{3L} \\ \nonumber
& & +\bar{U}_{3L}\gamma_\mu U_{3L}+\bar{D}_{3L}\gamma_\mu D_{3L}\\ \nonumber
& & +\sum_{\alpha=1}^3(-\bar{e}^-_{\alpha L}\gamma_\mu e^-_{\alpha L}-\bar{\nu}_{e \alpha L}\gamma_\mu \nu_{e \alpha L}\\ \label{zppcb}
& & +\bar{N}^0_{\alpha L}\gamma_\mu N^0_{\alpha L}
+\bar{E}^-_{\alpha L}\gamma_\mu E^-_{\alpha L}),
\end{eqnarray}

\begin{table*}
\caption{\label{tab:7}The $Z_1^\mu\longrightarrow \bar{f}f$ couplings for Model B.}
\begin{ruledtabular}
\begin{tabular}{lcc}
$f$ & $g^{(B)}(f)_{1V}$ & $g^{(B)}(f)_{1A}$ \\ \hline
$u_{1,2,3}$& $\cos\theta ({1\over 2}-{4S_W^2 \over 3})-
\frac{5\sin\theta}{6(C_{2W})^{1/2}}S_W^2$
& ${1\over 2} \cos\theta + \frac{\sin\theta}{2(C_{2W})^{1/2}}S_W^2$ \\ 
$d_{1,2,3}$ & $(-{1\over 2}+{2S_W^2\over 3})\cos\theta +\frac{\sin\theta}{6(C_{2W})^{1/2}}S^2_W$ 
& $-{1\over 2}\cos\theta - \frac{\sin\theta}{2(C_{2W})^{1/2}}S_W^2$ \\
$D_{1,2,3}$ & ${2S_W^2\over 3}\cos\theta +\frac{\sin\theta}{2(C_{2W})^{1/2}}({7 S_W^2 \over 3}-1)$ &
$-\frac{\sin\theta}{2(C_{2W})^{1/2}} C_W^2$ \\  
$U_{1,2,3}$ & $-{4S_W^2\over 3}\cos\theta-\frac{\sin\theta}{2(C_{2W})^{1/2}}({11 S_W^2 \over 3}-1)$ &
$\frac{\sin\theta}{2(C_{2W})^{1/2}} C_W^2$ \\
$e^-_{1,2,3}$& $\cos\theta (-{1\over 2}+2S_W^2)+ 
\frac{5\sin\theta}{2(C_{2W})^{1/2}}S_W^2 $ & 
$ -{\cos\theta\over 2} -\frac{\sin\theta}{2(C_{2W})^{1/2}}S_W^2$\\
$\nu_{1,2,3}$ &
${1\over 2}\cos\theta+ \frac{\sin\theta}{2(C_{2W})^{1/2}}S_W^2$ &
$ {1\over 2}\cos\theta+ \frac{\sin\theta}{2(C_{2W})^{1/2}}S_W^2$ \\ 
$N^0_{1,2,3}$ & $\frac{\sin\theta}{2(C_{2W})^{1/2}}C_W^2$ &
$\frac{\sin\theta}{2(C_{2W})^{1/2}}C_W^2$ \\ 
$E^-_{1,2,3}$ &
$2 S^2_W \cos\theta +\frac{\sin\theta}{(C_{2W})^{1/2}}(2-{5\over 2}C_W^2)$ & $-\frac{\sin\theta}{2(C_{2W})^{1/2}}C_W^2$ \\
\end{tabular}
\end{ruledtabular}
\end{table*}

\begin{table*}
\caption{\label{tab:8}The $Z_2^\mu\longrightarrow \bar{f}f$ couplings for Model B.}
\begin{ruledtabular}
\begin{tabular}{lcc}
$f$ & $g^{(B)}(f)_{2V}$ & $g^{(B)}(f)_{2A}$ \\ \hline
$u_{1,2,3}$& $-\sin\theta ({1\over 2}-{4S_W^2 \over 3})-
\frac{5\cos\theta}{6(C_{2W})^{1/2}}S_W^2$
& $-{1\over 2} \sin\theta + \frac{\cos\theta}{2(C_{2W})^{1/2}}S_W^2$ \\ 
$d_{1,2,3}$ & $({1\over 2}-{2S_W^2\over 3})\sin\theta +\frac{\cos\theta}{6(C_{2W})^{1/2}}S^2_W$ 
& ${1\over 2}\sin\theta - \frac{\cos\theta}{2(C_{2W})^{1/2}}S_W^2$ \\
$D_{1,2,3}$ & $-{2S_W^2\over 3}\sin\theta +\frac{\cos\theta}{2(C_{2W})^{1/2}}({7 S_W^2 \over 3}-1)$ &
$-\frac{\cos\theta}{2(C_{2W})^{1/2}} C_W^2$ \\  
$U_{1,2,3}$ & ${4S_W^2\over 3}\sin\theta-\frac{\cos\theta}{2(C_{2W})^{1/2}}({11 S_W^2 \over 3}-1)$ &
$\frac{\cos\theta}{2(C_{2W})^{1/2}} C_W^2$ \\
$e^-_{1,2,3}$& $\sin\theta ({1\over 2}-2S_W^2)+ 
\frac{5\cos\theta}{2(C_{2W})^{1/2}}S_W^2 $ & 
$ {\sin\theta\over 2} -\frac{\cos\theta}{2(C_{2W})^{1/2}}S_W^2$\\
$\nu_{1,2,3}$ &
$-{1\over 2}\sin\theta+ \frac{\cos\theta}{2(C_{2W})^{1/2}}S_W^2$ &
$-{1\over 2}\sin\theta+ \frac{\cos\theta}{2(C_{2W})^{1/2}}S_W^2$ \\ 
$N^0_{1,2,3}$ & $\frac{\cos\theta}{2(C_{2W})^{1/2}}C_W^2$ &
$\frac{\cos\theta}{2(C_{2W})^{1/2}}C_W^2$ \\ 
$E^-_{1,2,3}$ &
$-2 S^2_W \sin\theta +\frac{\cos\theta}{(C_{2W})^{1/2}}(2-{5\over 2}C_W^2)$ & $-\frac{\cos\theta}{2(C_{2W})^{1/2}}C_W^2$ \\
\end{tabular}
\end{ruledtabular}
\end{table*}

Note that $J_\mu(Z'')$ is a pure left-handed current and that, notwithstanding the extra neutral gauge boson $Z''_\mu$ does not mix neither with $Z_\mu$ nor with $Z'_\mu$ (for the particular case $V\simeq V^\prime$ and $v\simeq v^\prime$), it still couples nondiagonally to ordinary fermions. Thus, at low energy, we have tree-level FCNC transmitted by $Z''_\mu$. 

The left-handed currents in (\ref{zcb}) and (\ref{zpcb}) are
\begin{eqnarray} \nonumber
J_{\mu,L}(Z)&=&{1\over 2}[\sum_{i=1}^2(\bar{u}_{iL}\gamma_\mu u_{iL}
-\bar{d}_{iL}\gamma_\mu d_{iL})\\ \nonumber
& & -(\bar{d}_{3L}\gamma_\mu d_{3L}- \bar{u}_{3L}\gamma_\mu u_{3L})\\ 
\nonumber
& & -\sum_{\alpha=1}^3(\bar{e}^-_{\alpha L}\gamma_\mu e^-_{\alpha L}
-\bar{\nu}_{e\alpha L}\gamma_\mu \nu_{e\alpha L})] \\ \label{lhzcb}
 &=&\sum_f T_{4f}\bar{f}_L\gamma_\mu f_L ,
\end{eqnarray}
\begin{eqnarray}\nonumber
J_{\mu,L}(Z')&=&(2T_{W})^{-1}\lbrace\sum_{i=1}^2\lbrack T^2_W(\bar{u}_{iL}\gamma_\mu u_{iL}
-\bar{d}_{iL}\gamma_\mu d_{iL}) \\ \nonumber
& & -\bar{D}_{iL}\gamma_\mu D_{iL}+\bar{U}_{iL}\gamma_\mu U_{iL}\rbrack \\ \nonumber 
& & -T^2_W(\bar{d}_{3L}\gamma_\mu d_{3L}-\bar{u}_{3L}\gamma_\mu u_{3L}) \\ \nonumber
& & +\bar{U}_{3L}\gamma_\mu U_{3L}
-\bar{D}_{3L}\gamma_\mu D_{3L}\\ \nonumber 
& & +\sum_{\alpha=1}^3\lbrack -T^2_W(\bar{e}^-_{\alpha L}\gamma_\mu e^-_{\alpha L}- \bar{\nu}_{\alpha L}\gamma_\mu \nu_{\alpha L}) \\ \nonumber
& & +\bar{N}^0_{\alpha L}\gamma_\mu N^0_{\alpha L}-\bar{E}^-_{\alpha L}\gamma_\mu E^-_{\alpha L}\rbrack \rbrace
\\ \label{lhzpcb}
&=&\sum_f
T'_{4f}\bar{f}_L\gamma_\mu f_L,
\end{eqnarray}
where again $T_{4f}=Dg(1/2,-1/2,0,0)$ is the third component of the weak isospin, while $T'_{4f}=(1/2T_W)Dg(T^2_W,$ $-T^2_W,-1,1)$= $T_W\lambda_3/2 +(1/T_W)(\lambda_8/(2\sqrt{3})-\lambda_{15}/\sqrt{6})$ is a convenient $4\times 4$ diagonal matrix, both of them acting on the representation 4 of $SU(4)_L$. Since $J_\mu(Z)$ is the generalization of the neutral current present in the SM, we can identify $Z_\mu$ as the neutral gauge boson of the SM. Notice from Eq.~(\ref{lhzpcb}) that the left-handed couplings of fermions to $Z^\prime$ are flavor-diagonal so, there are not tree-level FCNC transmitted by the $Z^\prime$ gauge boson in this model.

The couplings $g^{(B)}_{iV},\; g^{(B)}_{iA}$ ($i=1,2$) of the fermion fields to the mass eigenstates $Z_1^\mu$ and $Z_2^\mu$ for Model B are listed in Tables~\ref{tab:7} and \ref{tab:8} from which we see that, unlike Model A, these couplings are family universal. This is a direct consequence of the fact that, according Table~\ref{tab:5}, the three families of quarks have the same hypercharge $X$.
 
\section{\label{sec:sec3}Constraints on the parameters}

\subsection{\label{sec:sub3a}Constraints on the parameter space ($\theta-M_{Z_2}$) from Z-pole observables and APV data}
We start constrainig the parameter space ($\theta-M_{Z_2}$) for Models A and B by using electroweak observables measured at the $Z$-pole from the CERN $e^+e^-$ collider (LEP), SLAC Linear Collider, and atomic parity violation data which are given in Table~\ref{tab:9}. 

The partial decay width for $Z^{\mu}_1\rightarrow f\bar{f}$ is given, in the on-shell scheme, by \cite{pdg,pich}
 \begin{eqnarray}\nonumber
\Gamma(Z^{\mu}_1\rightarrow f\bar{f})&=&\frac{N_C G_F
M_{Z_1}^3}{6\pi\sqrt{2}}\rho_f \Big\{\frac{3\beta-\beta^3}{2}
[g(f)_{1V}]^2 \\ \nonumber
& + & \beta^3[g(f)_{1A}]^2 \Big\}(1+\delta_f)R_{EW}R_{QCD}, \\ \label{ancho}
\end{eqnarray}
\noindent 
where $f$ is an ordinary SM fermion, $Z^\mu_1$ is the physical gauge boson
observed at LEP, $N_C=1$ for leptons while for quarks
$N_C=3(1+\alpha_s/\pi + 1.405\alpha_s^2/\pi^2 - 12.77\alpha_s^3/\pi^3)$,
where the 3 is due to color and the factor in parentheses represents the
universal part of the QCD corrections for massless quarks. $R_{EW}$ are electroweak corrections which include the leading order QED corrections given by $R_{\mathrm{QED}}$ $=1+3\alpha q^2_f/(4\pi)$. $R_{\mathrm{QCD}}$ are further QCD corrections, and $\beta=\sqrt{1-4 m_f^2/M_{Z_1}^2}$ is a kinematic factor which can be taken equal to $1$ for all the SM fermions except for the bottom quark. 
The factor $\delta_f$ contains the one loop vertex
contribution which is negligible for all fermion fields except for the 
bottom quark for which the contribution coming from the top quark at the 
one loop vertex radiative correction is parametrized as $\delta_b\approx 
10^{-2} [-m_t^2/(2 M_{Z_1}^2)+1/5]$. The parameter $\rho_f$ 
is written as
\begin{equation}
\rho_f = 1+\rho_t + \rho_V,
\end{equation} 
where
\begin{equation} 
\rho_t \approx 3G_F m_t^2/(8\pi^2\sqrt{2}),
\end{equation}
with $m_t$ being the top quark pole mass. $\rho_V$ is the tree-level contribution due to the $(Z_{\mu} - Z'_{\mu})$ mixing which can be parametrized as
\begin{equation}\label{rov}
\rho_V\approx (M_{Z_2}^2/M_{Z_1}^2-1)\sin^2\theta. 
\end{equation}
Universal electroweak corrections are included in $\rho_t$ and in the couplings $g^{(k)}(f)_{1V}$ and $g^{(k)}(f)_{1A}$ ($k= A,B$) of the physical $Z_1^\mu$ field with ordinary fermions which are written in terms of the effective Weinberg angle $\bar{S}^2_W = (1+\rho_t/T^2_W)S^2_W$.

\begin{table}
\caption{\label{tab:9}Experimental data and SM values for some observables used for the $\chi^2$ fit.}
\begin{ruledtabular}
\begin{tabular}{lcl}
& Experimental results & SM \\ \hline
$\Gamma_Z$[GeV]  & $2.4952 \pm 0.0023$  &  $2.4968 \pm 0.0010$  \\   
$\Gamma(\mathrm{had})$ [GeV]  & $1.7444 \pm 0.0020$ & $1.7434 \pm 0.0010$ \\ 
$\Gamma(l^+l^-)$ [MeV] & $83.984\pm 0.086$ & $83.988 \pm 0.016$ \\
$R_e$ & $20.804\pm 0.050$ & $20.758\pm 0.011$ \\ 
$R_\mu$ & $20.785\pm 0.033$ & $20.758\pm 0.011$ \\ 
$R_\tau$ & $20.764\pm 0.045$ & $20.803\pm 0.011$ \\ 
$R_b$ & $0.21629\pm 0.00066$ & $0.21584\pm 0.00006$ \\ 
$R_c$ & $0.1721\pm 0.0030$ & $0.17228\pm 0.00004$ \\
$A^{(0,e)}_{FB}$ & $0.0145\pm 0.0025$ & $0.01627\pm 0.00023$ \\
$A^{(0,\mu)}_{FB}$ & $0.0169\pm 0.0013$ & $$ \\
$A^{(0,\tau)}_{FB}$ & $0.0188\pm 0.0017$ & $$ \\
$A^{(0,b)}_{FB}$ & $0.0992\pm 0.0016$ & $0.1033\pm 0.0007$ \\
$A^{(0,c)}_{FB}$ & $0.0707\pm 0.0035$ & $0.0738\pm 0.0006$ \\
$A^{(0,s)}_{FB}$ & $0.0976\pm 0.0114$ & $0.1034\pm 0.0007$ \\
$A_e$ & $0.15138\pm 0.00216$ & $0.1473\pm 0.0011$ \\
$A_\mu$ & $0.142\pm 0.015$ & $$ \\
$A_\tau$ & $0.136\pm 0.015$ & $$ \\
$A_b$ & $0.923\pm 0.020$ & $0.9348\pm 0.0001$ \\
$A_c$ & $0.670\pm 0.027$ & $0.6679\pm 0.0005$ \\
$A_s$ & $0.895\pm 0.091$ & $0.9357\pm 0.0001$ \\
$Q_W^{Cs}$ & $-72.62\pm 0.46$ & $-73.16\pm 0.03$ \\
$M_{Z_{1}}$[GeV] & $ 91.1876 \pm 0.0021 $ & $ 91.1874 \pm 0.0021 $ \\
$m_t$[GeV] & $170.9 \pm 1.8 \pm 0.6$ & $ 171.1 \pm 1.9 $ \\
\end{tabular}
\end{ruledtabular}
\end{table}

The ratios of partial widths are defined as 
\begin{equation}\label{rl}
R_l\equiv \frac{\Gamma_Z(\mathrm{had})}{\Gamma(l^+l^-)}\quad \mbox{for}\quad l=e,\mu,\tau,
\end{equation} 
and 
\begin{equation}\label{reta}
R_\eta\equiv \frac{\Gamma_\eta}{\Gamma_Z(\mathrm{had})}\quad \mbox{for}\quad \eta=b,c.
\end{equation} 

We shall use the experimental values:
$\alpha_s(M_{Z_1})=0.1198$, $\alpha(M_{Z_1})^{-1}=127.918$, and $\sin^2\theta_W=0.2231$. For the bottom quark mass we use the running mass $\widehat{m}_b(M_{Z_1})$ at the $Z_1$ scale in the $\widehat{MS}$ scheme \cite{borzu}
\begin{equation}\nonumber
\widehat{m}_b(M_{Z_1})=m_b\Big[ 1+\frac{\alpha_s(M_{Z_1})}{\pi}\Big( \mbox{ln}\frac{m^2_b}{M^2_{Z_1}}-\frac{4}{3}\Big) \Big], \label{bottom}
\end{equation} 
where $m_b \approx 4.25$ GeV is the pole mass.

The forward-backward asymmetries at the Z-pole are given by
\begin{eqnarray}\nonumber
A^{(0,f)}_{FB}=\frac{3}{4}A_eA_f,& \quad{\mbox{where}}\quad &A_f=\frac{2g(f)_{1V}g(f)_{1A}}{g(f)^{2}_{1V}+g(f)^{2}_{1A}}. \\ \label{fbasym}
\end{eqnarray}

The effective weak charge in atomic parity violation, $Q_W$, can be 
expressed as 
\begin{equation}
Q_W=-2\left[(2Z+N)c_{1u}+(Z+2N)c_{1d}\right], 
\end{equation}
\noindent
where $Z$ is the number of protons and $N$ the number of neutrons in the atomic nucleus, and $c_{1q}=2g(e)_{1A}g(q)_{1V}$ ($q=u,d$). Notice that the values of the coefficients $c_{1q}$ depend, for each model, on the assignment of quark gauge eigenstates into mass eigenstates.

The theoretical value for $Q_W$ for the Cesium atom is given by \cite{guena} $Q_W(^{133}_{55}Cs)=-73.17\pm0.03 + \Delta Q_W$, where the contribution of new physics is included in $\Delta Q_W$ and can be written as \cite{durkin,alta}

\begin{equation}\label{DQ} 
\Delta 
Q_W=\left[\left(1+4\frac{S^4_W}{1-2S^2_W}\right)Z-N\right]\rho_V
+\Delta Q^\prime_W,
\end{equation}
where
\begin{eqnarray}\nonumber
\Delta Q^\prime_W &=& 16\lbrack (2Z+N)(g(e)_{1A}g(u)_{2V}+g(e)_{2A}g(u)_{1V}) \\ \nonumber
&+&(Z+2N)(g(e)_{1A}g(d)_{2V}+g(e)_{2A}g(d)_{1V})\rbrack \sin\theta \\ \nonumber
&-& 16\lbrack(2Z+N)g(e)_{2A}g(u)_{2V} \\
& & +(Z+2N)g(e)_{2A}g(d)_{2V}\rbrack
\frac{M^2_{Z_1}}{M^2_{Z_2}}. \label{dqprima} 
\end{eqnarray}

The term $\Delta Q^\prime_W$ is model dependent. In particular, it is a function of the couplings $g(q)_{2V}$ and $g(q)_{2A}$ ($q=u,d$) of the first family of quarks to the new neutral gauge boson $Z_2$. So, the new physics in $\Delta Q^\prime_W$ depends on which family of quarks transform differently under the gauge group. For each assignment in Table~\ref{tab:2}, we get the following values for $\Delta Q^{\prime}_W$ in Model A
 
\begin{eqnarray}\nonumber
\mbox{Assignments A1, A2:}& & \\ \nonumber
\Delta Q_W^\prime=&(10.63 Z + 6.99 N) \sin\theta & \\
&+ (4.94 Z + 4.18 N)\frac{M^2_{Z_1}}{M^2_{Z_2}}, & \label{dqpA1A2}
\end{eqnarray}
\begin{eqnarray}\nonumber
\mbox{Assignment A3:}& & \\ \nonumber
\Delta Q_W^\prime=&(-5.51 Z - 9.15 N) \sin\theta & \\
&+ (-2.66 Z - 3.42 N)\frac{M^2_{Z_1}}{M^2_{Z_2}}, & \label{dqpA3}
\end{eqnarray}
while in Model B, for which there is not family dependence in the fermion couplings to $Z_1$ and $Z_2$, we obtain
\begin{equation}\label{dqpB}
\Delta Q_W^\prime=(3.66 Z + 2.51 N) \sin\theta + (-1.18 Z - 0.39 N)
\frac{M^2_{Z_1}}{M^2_{Z_2}}\; .
\end{equation}

The discrepancy between the SM and the experimental data for $\Delta Q_W$ 
is given by \cite{guena}

\begin{equation}
\Delta Q_W=Q^{exp}_W-Q^{SM}_W=0.45\pm 0.48,
\end{equation}
which is $1.1\; \sigma$ away from the SM predictions.

Using the couplings $g^{(k)}(f)_{iV}$ and $g^{(k)}(f)_{iV}$ ($k=A,B$, $i=1,2$) for each model, introducing the expressions for $Z$-pole observables in Eqs.~(\ref{ancho}), (\ref{rl}), (\ref{reta}) and (\ref{fbasym}), and with $\Delta Q_W$ in terms of new physics in Eq.~(\ref{DQ}), and using experimental data from LEP, SLAC Linear Collider and atomic parity violation (see Table~\ref{tab:9}), we do a $\chi^2$ fit and find the best allowed region in the $(\theta-M_{Z_2})$ plane at $95\%$ confidence level (C.L.).
  
\begin{figure} 
\begin{center}
\resizebox{0.98\textwidth}{!}{
\includegraphics{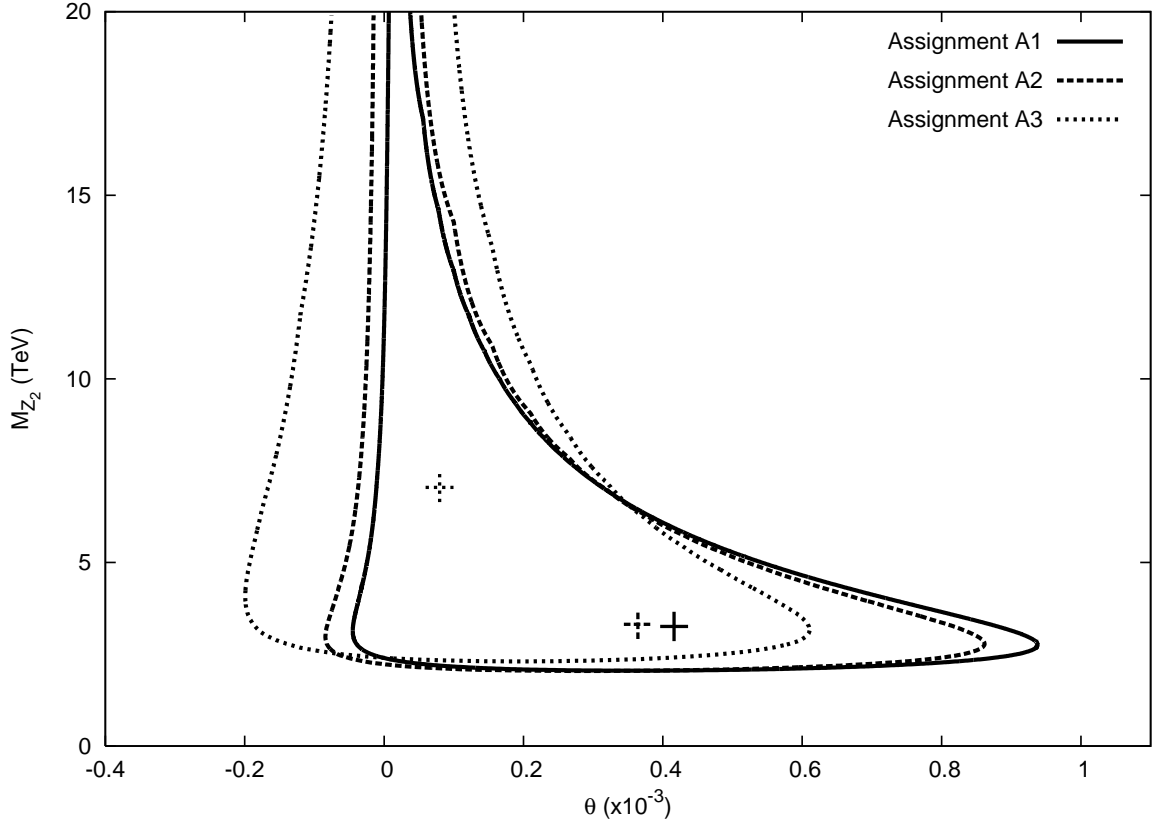}
}
\caption{\label{fig:1}Contour plot displaying the allowed regions for
$\theta$ vs $M_{Z_2}$ at $95\%$ C.L. for Model A and for the three assignments in Table 2. The crosses locate the best fit values.}
\end{center}
\end{figure} 

In Fig.~\ref{fig:1} we display this region for each assignment in Model A. From this figure we get the family-dependent constraints 
\begin{eqnarray}\nonumber
&\mbox{Assignment A1:} \\ \label{boundA1}
&-0.00004 \leq \theta \leq 0.00094, \;\;\; 2.03\; {\mbox TeV} \leq M_{Z_2},
\end{eqnarray}
\begin{eqnarray}\nonumber
& \mbox{Assignment A2:} \\ \label{boundA2}
&-0.00008 \leq \theta \leq 0.00087, \;\;\; 2.03\; {\mbox TeV} \leq M_{Z_2}, \end{eqnarray}
\begin{eqnarray}\nonumber
&\mbox{Assignment A3:} \\ \label{boundA3}
&-0.00019 \leq \theta \leq 0.00062, \;\;\; 2.31\; {\mbox TeV} \leq M_{Z_2}, 
\end{eqnarray}

\begin{figure} 
\begin{center}
\resizebox{0.98\textwidth}{!}{
\includegraphics{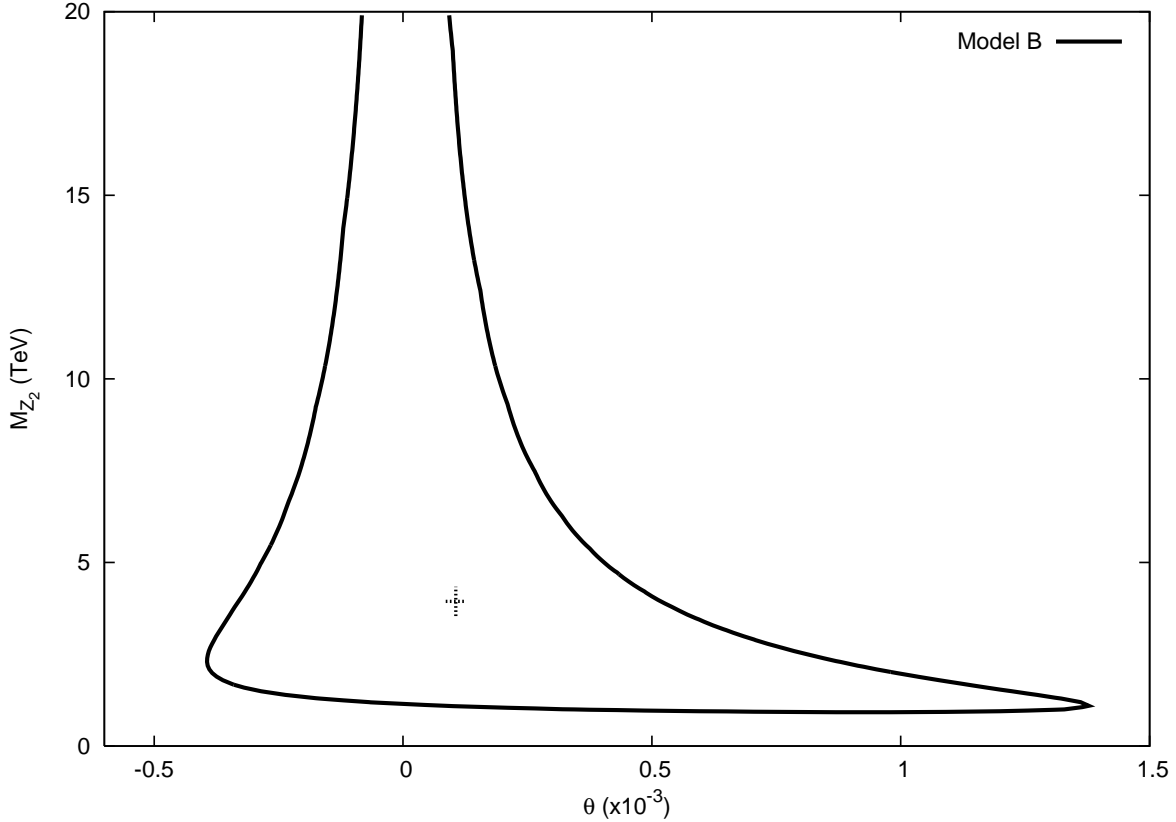}
}
\caption{\label{fig:2}Contour plot displaying the allowed region for
$\theta$ vs $M_{Z_2}$ at $95\%$ C.L. for Model B. The cross locates the best fit values.}
\end{center}
\end{figure}

For Model B the allowed region is shown in Fig.~\ref{fig:2}, which produces the family-independent constraints \cite{correction}
\begin{equation} 
-0.00039 \leq \theta \leq 0.00139, \;\;\; 0.89\; {\mbox TeV} \leq M_{Z_2}. \label{boundsB}
\end{equation}

For Model A the allowed region in the $(\theta-M_{Z_2})$ plane depends on which generation of quarks transforms differently under $SU(4)_L\otimes U(1)_X$. So, for this model the phenomenology depends crucially on the family assignments in Table~\ref{tab:2}. For example, Eqs.~(\ref{boundA1}) and (\ref{boundA2}) show that the second family or the third one must be different in order to keep the bound on $M_{Z_2}$ as low as possible. 

Unlike Model A, for Model B the allowed region in the $(\theta-M_{Z_2})$ plane does not depend on the quark family assignments in Table~\ref{tab:6}. This is a consequence of the family-universal character of the couplings in Tables~\ref{tab:7} and \ref{tab:8}, which in turn comes from the fact that in this model the three quark families have the same hypercharge $X$.

It must be stressed that for both models the constraints on $M_{Z_2}$ are compatible with the bound obtained in $p\bar{p}$ collisions at the Fermilab Tevatron \cite{abe}. 

\subsection{\label{sec:sub3b}Bounds from FCNC processes} 
The discrete $Z_2$ symmetries introduced in Models A and B not only produce a simple mass splitting between ordinary and exotic fermions, but also avoid unitarity violation of the CKM matrix arising from their mixing. Since each flavor couples to more than one Higgs 4-plet, FCNC coming from the scalar sector are also present. Because this last contribution depends on a large number of arbitrary parameters, is not very useful in order to constrain the model and we will ignore it. Because the right-handed quarks transform as $SU(4)_L$ singlets, they are generation universal and, consequently, they couple diagonally to the neutral gauge bosons. However, there is an additional source of FCNC due to the fact that anomaly cancellation among the families forces us to have one family of left-handed quarks in the weak basis transforming differently from the other two.

Regarding the left-handed interaction of quarks, Eqs.~(\ref{lhzca}) and (\ref{lhzcb}) show that the couplings to the $Z$ boson remain flavor conserving for both models. Then, neglecting the scalar contribution and because of the $Z_2$ symmetry, the only source of FCNC is in the left-handed interactions of ordinary quarks with the new neutral gauge boson $Z^\prime$ in Model A and with $Z^{\prime \prime}$ in Model B. For their study we will follow the analysis presented in Refs.~\cite{liu,jars} where bounds coming from neutral meson mixing are obtained in the context of the so-called ``minimal 3-3-1 model", which is a model with extra quarks having exotic electric charges.

\subsubsection{\label{sec:sub3b1}Model A}

With $b=c=1$ and from the charge operator in Eq.~(\ref{ch}), the $Y$ hypercharge of the SM is given by $Y/2=T_{8L}/\sqrt{3}+T_{15L}/\sqrt{6}+X$.
In terms of $Y$, the couplings of $Z^\prime$ to left-handed quarks in Eqs.~(\ref{zpca}) and (\ref{lhzpca}), can be written in a more convenient fashion for 4-plets as

\begin{equation}\label{Lzprime}
{\cal L}(Z^\prime)=\frac{g}{2C_W} \frac{1}{\sqrt{6}\sqrt{2-3S^2_W}}Z^\prime_\mu J^\mu(Z^\prime),
\end{equation}
with

\begin{equation}
J^\mu(Z^\prime)=\sum_f \bar{f}\gamma^\mu[\sqrt{6}S^2_WY-4C^2_WT_L]P_Lf.
\end{equation}
where $P_L$ is the left-handed projector and $T_L=\sqrt{2}T_{8L}+T_{15L}$.

The value of the operator $T_L$ is not the same for 4-plets than for $4^*$-plets. Then, the flavor changing interaction can be written, for ordinary up- and down-type quarks $q^\prime$ in the weak basis, as
\begin{equation}\label{zfc}
J^\mu(Z^\prime)_{\mathrm{FCNC}}=-4C^2_W\sum_{q^\prime} \bar{q^\prime}\gamma^\mu[T_L(4)-T_L(4^*)]P_Lq^\prime.
\end{equation}

Using (\ref{Lzprime}) and (\ref{zfc}) we have
\begin{eqnarray}\nonumber
{\cal L}(Z^\prime)_{\mathrm{FCNC}}&=&-\frac{gC^2_W}{\sqrt{2-3S^2_W}}(\sin\theta Z^\mu_1+\cos\theta Z^\mu_2) \\ \label{Lfcnc}
& & \times\sum_{q^\prime} \bar{q^\prime}\gamma_\mu P_L q^\prime,
\end{eqnarray}
where, by using Eq.~(\ref{mixa}), we have included the mass eigenstates $Z_1$ and $Z_2$. 

In order to consider constraints coming from experimental data in the $K^0-\bar{K}^0$, $B^0_d-\bar{B}^0_d$, $B^0_s-\bar{B}^0_s$ and $D^0-\bar{D}^0$ systems, we first notice that the submatrices $M^u_{3\times 3}$ and $M^d_{3\times 3}$ in Eqs.~(\ref{mupa}) and (\ref{mdowna}) are diagonalized by biunitary transformations $U_{L,R}$ and $V_{L,R}$, respectively, with $V_{\mathrm{CKM}}= U^{\dagger}_LV_L$ being the CKM mixing matrix. Then, in terms of mass eigenstates, Eq.~(\ref{Lfcnc}) produces the following effective Hamiltonian for the tree-level neutral meson mixing interactions 
\begin{eqnarray}\nonumber
{\cal H}^{(\alpha,\beta)}_{\mathrm{eff}}&=&\frac{2\sqrt{2}G_F C^4_W \cos^2\theta}{2-3S^2_W}(V^*_{Lj\alpha}V_{Lj\beta})^2 \\ \label{heff}
& & \times \biggl(\frac{M^2_{Z_1}}{M^2_{Z_2}}+\tan^2\theta\biggr)[\bar{\alpha}\gamma_\mu P_L\beta]^2,
\end{eqnarray}
where $(\alpha,\beta)$ must be replaced by $(d,s)$, $(d,b)$, $(s,b)$ and $(u,c)$ for the $K^0-\bar{K}^0$, $B^0_d-\bar{B}^0_d$, $B^0_s-\bar{B}^0_s$ and $D^0-\bar{D}^0$ systems, respectively, and $V_L$ must be replaced by $U_L$ for the neutral $D^0-\bar{D}^0$ system. The family index $j= 1,2,3$ refers to the family of quarks to be singled out as transforming differently.

The effective Hamiltonian gives the following contribution to the mass difference $\Delta m_K$ 
\begin{eqnarray}\nonumber
\frac{\Delta m_K}{m_K}&=&\frac{4\sqrt{2}G_F C^4_W \cos^2\theta}{3(2-3S^2_W)}\mbox{Re} [(V^*_{Ljd}V_{Ljs})^2] \\ \label{mdifK}
& & \times \;\eta_K \biggl(\frac{M^2_{Z_1}}{M^2_{Z_2}}+\tan^2\theta\biggr)B_Kf^2_K,
\end{eqnarray}
while for the $B^0_d-\bar{B}^0_d$, $B^0_s-\bar{B}^0_s$ and $D^0-\bar{D}^0$ systems, we have
\begin{eqnarray}\nonumber
\frac{\Delta m_B}{m_B}&=&\frac{4\sqrt{2}G_F C^4_W \cos^2\theta}{3(2-3S^2_W)}\vert V^*_{Lj\alpha}V_{Lj\beta}\vert^2 \\ \label{mdifB}
& & \times \;\eta_B \biggl(\frac{M^2_{Z_1}}{M^2_{Z_2}}+\tan^2\theta\biggr)B_Bf^2_B,
\end{eqnarray}

\begin{eqnarray}\nonumber
\frac{\Delta m_D}{m_D}&=&\frac{4\sqrt{2}G_F C^4_W \cos^2\theta}{3(2-3S^2_W)}\vert U^*_{Lju}U_{Ljc}\vert^2 \\ \label{mdifD}
& & \times \;\eta_D \biggl(\frac{M^2_{Z_1}}{M^2_{Z_2}}+\tan^2\theta\biggr)B_Df^2_D,
\end{eqnarray}
\noindent
where $B$ stands for $B_d$ or $B_s$. $B_m$ and $f_m$ ($m=K, B_d, B_s,$ $D$) are the bag parameter and decay constant of the corresponding neutral meson, respectively. The $\eta$'s are QCD corrections which, at leading order, can be taken as equal to the ones of the SM \cite{blanke}, that is: $\eta_K\simeq \eta_D\simeq 0.57$, $\eta_{B_d} = \eta_{B_s}\simeq 0.55$ \cite{gilman}. 

Several sources, different from the tree-level $Z_2$ exchange, may contribute to the mass differences and it is not possible to disentangle the $Z_2$ contribution from other effects. because of this, several authors consider reasonable to assume that the $Z_2$ exchange contribution must not be larger than the experimental values \cite{liu}. In this work we will assume that this is the case. We must notice, however, that more conservative but rather arbitrary criteria have been used by other authors \cite{blanke}.

\begin{table}
\caption{\label{tab:10}Values of the experimental and theoretical quantities used as input parameters for FCNC processes.}
\begin{ruledtabular}
\begin{tabular}{llc}
& Value & Reference \\ \hline
$\Delta m_K$ [GeV] & $3.483(6)\times 10^{-15}$ & \cite{pdg} \\   
$m_{K^0}$ [MeV] & $497.65(2)$ & \cite{pdg} \\ 
$f_K\sqrt{B_K}$ [MeV] & $143(7)$ & \cite{hasi} \\ \hline
$\Delta m_{B_d}$ [$\mbox{ps}^{-1}$] & $0.508(4)$ & \cite{pdg} \\ 
$m_{B_d}$ [GeV] & $5.2794(5)$ & \cite{pdg} \\ 
$f_{B_d}\sqrt{B_{B_d}}$ [MeV] & $214(38)$ & \cite{hasi} \\ \hline
$\Delta m_{B_s}$ [$\mbox{ps}^{-1}$] & $17.77(12)$ & \cite{abuaba} \\ 
$m_{B_s}$ [GeV] & $5.370(2)$ & \cite{pdg} \\ 
$f_{B_s}\sqrt{B_{B_s}}$ [MeV] & $262(35)$ & \cite{hasi} \\ \hline
$\Delta m_D$ [$\mbox{ps}^{-1}$] & $11.7(6.8)\times 10^{-3}$ & \cite{ciu} \\   
$m_{D^0}$ [GeV] & $1.8645(4)$ & \cite{pdg} \\
$f_D\sqrt{B_D}$ [MeV] & $241(24)$ & \cite{arlin} \\ \hline
$m_u(M_Z)$ [MeV] & $2.33^{+0.42}_{-0.45}$ & $$ \\   
$m_c(M_Z)$ [MeV] & $677^{+56}_{-61}$ & $$ \\ 
$m_t(M_Z)$ [GeV] & $181\pm 13$ & $$ \\
$m_d(M_Z)$ [MeV] & $4.69^{+0.60}_{-0.66}$ & $$ \\
$m_s(M_Z)$ [MeV] & $93.4^{+11.8}_{-13.0}$ & $$ \\
$m_b(M_Z)$ [GeV] & $3.00\pm 0.11$ & \cite{fusa} \\
\end{tabular}
\end{ruledtabular}
\end{table}
 
Since the complex numbers $V_{Lij}$ and $U_{Lij}$ cannot be estimated from the present experimental data, and in order to compare with the bounds obtained in the previous section, we assume the Fritzsch ansatz for the quark mixing matrix \cite{frit}, which implies (for $i\leq j$) $V_{Lij}=\sqrt{m_i/m_j}$, and similarly for $U_L$ \cite{cheng} ({\it CP} violating phases in the mixing matrices will not be considered here). 
To obtain bounds on $M_{Z_2}$ from Eqs.~(\ref{mdifK}), (\ref{mdifB}) and (\ref{mdifD}), we use updated experimental and theoretical values for the input parameters as shown in Table~\ref{tab:10}, where the quark masses are given at Z-pole. For Model A and for the three different assignments in Table~\ref{tab:2}, the results are
\begin{eqnarray}\nonumber
\mbox{Assignment A1:} & \\ \nonumber
K^0-\bar{K}^0: & M_{Z_2}>3.66\; \mbox{TeV}, \\ \nonumber
B^0_d-\bar{B}^0_d: & M_{Z_2}>11.34 \;\mbox{TeV}, \\ \nonumber
B^0_s-\bar{B}^0_s: & M_{Z_2}>10.56\; \mbox{TeV}, \\ \label{lowerbA1}
D^0-\bar{D}^0: & M_{Z_2}>0.18\; \mbox{TeV}.  
\end{eqnarray}
\begin{eqnarray}\nonumber
\mbox{Assignment A2:} & \\ \nonumber
K^0-\bar{K}^0: & M_{Z_2}>118.63\; \mbox{TeV}, \\ \nonumber
B^0_d-\bar{B}^0_d: & M_{Z_2}>11.34\;\mbox{TeV}, \\ \nonumber
B^0_s-\bar{B}^0_s: & M_{Z_2}>10.75\; \mbox{TeV}, \\ \label{lowerbA2}
D^0-\bar{D}^0: & M_{Z_2}>47.84\; \mbox{TeV}.  
\end{eqnarray}
\begin{eqnarray}\nonumber
\mbox{Assignment A3:} & \\ \nonumber
K^0-\bar{K}^0: & M_{Z_2}>118.63\; \mbox{TeV}, \\ \nonumber
B^0_d-\bar{B}^0_d: & M_{Z_2}>11.34 \;\mbox{TeV}, \\ \nonumber
B^0_s-\bar{B}^0_s: & M_{Z_2}>0.53\; \mbox{TeV}, \\ \label{lowerbA3}
D^0-\bar{D}^0: & M_{Z_2}>47.84\; \mbox{TeV}.  
\end{eqnarray}

This shows that the bounds on $M_{Z_2}$ coming from FCNC are also family dependent. For the assignment A1, the strongest constraint comes from the $B^0_d-\bar{B}^0_d$ system which reduces the parameter space $\theta-M_{Z_2}$ by raising the lower bound on $M_{Z_2}$ to a value larger than $11.34$ TeV, as compared with the bound in Eq.~(\ref{boundA1}). For the assignments A2 and A3, instead, the strongest constraint comes from the $K^0-\bar{K}^0$ system which raises this bound to a very large value greater than $118.63$ TeV, as compared with the bounds in Eqs.~(\ref{boundA2}) and (\ref{boundA3}). Since the assignment A1 singles out the third family to be different, this must be the case if we want to keep the lower bound on the $Z_2$ mass as low as possible.

\subsubsection{\label{sec:sub3b2}Model B}

As already stated, FCNC in Model B are present only in the Lagrangian for the neutral current $Z^{\mu\prime\prime}\equiv Z^{\mu}_3$ which can be written, for 4-plets, as
\begin{equation}\label{Lzpp}
{\cal L}(Z_3)=-\frac{g}{\sqrt{3}}Z^{\mu}_3 J_\mu(Z_3),
\end{equation}
with

\begin{equation}
J^\mu(Z_3)=\sum_f \bar{f}\gamma^\mu T_L P_Lf,
\end{equation}
where again $T_L=\sqrt{2}T_{8L}+T_{15L}$. Then, for ordinary up- and down-type quarks $q^\prime$, the flavor changing interaction is given by
\begin{equation}\label{zppfc}
J^\mu(Z_3)_{\mathrm{FCNC}}=\sum_{q^\prime} \bar{q^\prime}\gamma^\mu[T_L(4)-T_L(4^*)]P_Lq^\prime,
\end{equation}
so that
\begin{equation}\label{Lppfcnc}
{\cal L}(Z_3)_{\mathrm{FCNC}}=-\frac{g}{\sqrt{2}} Z^\mu_3 \sum_{q^\prime} \bar{q^\prime}\gamma_\mu P_L q^\prime.
\end{equation}

In terms of mass eigenstates, this Lagrangian produces the following effective Hamiltonian for the tree-level neutral meson mixing interactions
\begin{equation}\label{heffb}
{\cal H}^{(\alpha,\beta)}_{\mathrm{eff}}=\sqrt{2}G_F C^2_W (V^*_{Lj\alpha}V_{Lj\beta})^2 
\frac{M^2_{Z_1}}{M^2_{Z_3}}[\bar{\alpha}\gamma_\mu P_L\beta]^2,
\end{equation}
with the same meaning for $(\alpha,\beta)$ as before.

The contribution from ${\cal H}^{(\alpha,\beta)}_{\mathrm{eff}}$ to the mass differences in the $K^0-\bar{K}^0$, $B^0_d-\bar{B}^0_d$, $B^0_s-\bar{B}^0_s$ and $D^0-\bar{D}^0$ systems are given, respectively, by

\begin{eqnarray}\nonumber
\frac{\Delta m_K}{m_K}&=&\frac{2\sqrt{2}G_F C^2_W }{3}\mbox{Re} [(V^*_{Ljd}V_{Ljs})^2] \\ \label{mdifKb}
& & \times \;\eta_K \frac{M^2_{Z_1}}{M^2_{Z_3}}B_Kf^2_K,
\end{eqnarray}

\begin{eqnarray}\nonumber
\frac{\Delta m_B}{m_B}&=&\frac{2\sqrt{2}G_F C^2_W }{3}\vert V^*_{Lj\alpha}V_{Lj\beta}\vert^2 \\ \label{mdifBb}
& & \times \;\eta_B \frac{M^2_{Z_1}}{M^2_{Z_3}}B_Bf^2_B,
\end{eqnarray}

\begin{eqnarray}\nonumber
\frac{\Delta m_D}{m_D}&=&\frac{2\sqrt{2}G_F C^2_W }{3}\vert U^*_{Lju}U_{Ljc}\vert^2 \\ \label{mdifDb}
& & \times \;\eta_D \frac{M^2_{Z_1}}{M^2_{Z_3}}B_Df^2_D,
\end{eqnarray}
\noindent
where again $B$ stands for $B_d$ or $B_s$.

Assuming the same contribution to the mass differences from the $Z_3$ exchange as from the $Z_2$ exchange in Model A, and the Fritzsch ansatz for the quark mass matrices, the input parameters in Table~\ref{tab:10} give the following lower bounds on the $Z_3$ mass for the three assignments in Table~\ref{tab:6}
\begin{eqnarray}\nonumber
\mbox{Assignment B1:} & \\ \nonumber
K^0-\bar{K}^0: & M_{Z_3}>2.39\; \mbox{TeV}, \\ \nonumber
B^0_d-\bar{B}^0_d: & M_{Z_3}>6.61 \;\mbox{TeV}, \\ \nonumber
B^0_s-\bar{B}^0_s: & M_{Z_3}>6.16\; \mbox{TeV}, \\ \label{lowerbB1}
D^0-\bar{D}^0: & M_{Z_3}>0.16\; \mbox{TeV}.  
\end{eqnarray}
\begin{eqnarray}\nonumber
\mbox{Assignment B2:} & \\ \nonumber
K^0-\bar{K}^0: & M_{Z_3}>76.67\; \mbox{TeV}, \\ \nonumber
B^0_d-\bar{B}^0_d: & M_{Z_3}>6.61 \;\mbox{TeV}, \\ \nonumber
B^0_s-\bar{B}^0_s: & M_{Z_3}>6.16\; \mbox{TeV}, \\ \label{lowerbB2}
D^0-\bar{D}^0: & M_{Z_3}>44.04\; \mbox{TeV}.  
\end{eqnarray}
\begin{eqnarray}\nonumber
\mbox{Assignment B3:} & \\ \nonumber
K^0-\bar{K}^0: & M_{Z_3}>76.67\; \mbox{TeV}, \\ \nonumber
B^0_d-\bar{B}^0_d: & M_{Z_3}>6.61 \;\mbox{TeV}, \\ \nonumber
B^0_s-\bar{B}^0_s: & M_{Z_3}>0.31\; \mbox{TeV}, \\ \label{lowerbB3}
D^0-\bar{D}^0: & M_{Z_3}>44.04\; \mbox{TeV}.  
\end{eqnarray}

Evidently, these results show a lower bound on $M_{Z_3}$ depending on which family of quarks in Table~\ref{tab:5} is assigned to the $4^*$-plet. For the assignment B1, according to which the heaviest family of quarks transforms differently, the strongest constraint is imposed by the $B^0_d-\bar{B}^0_d$ system which gives the lower bound $M_{Z_3}>6.61$ TeV. For the assignments B2 and B3, which pick up the second and first family of quarks, respectively, the strongest constraint, $M_{Z_3}>76.67$ TeV, is provided by the $K^0-\bar{K}^0$ system. This shows that in Model B, like in Model A, the third family of quarks must transform differently in order to get, in the present case, the smallest lower bound on the $Z_3$ mass. Since $M^2_{Z_3}=(g^2_4/2)(V^2$ $+v^2)$, we also have a lower bound on the scale of breaking of the 3-4-1 symmetry.

Interestingly, the bound from the $B^0_d-\bar{B}^0_d$ system, both in Models A and B, is not family dependent. 

\section{Summary and Conclusions}
We have studied the impact of family dependence derived from quark family nonuniversality on the parameter space $\theta-M_{Z_2}$ of anomaly-free extensions of the SM based on the gauge group $SU(3)_c\otimes SU(4)_L\otimes U(1)_X$, which do not contain exotic electric charges. This last constraint picks up two classes of models characterized, respectively, by the values $b=c=1$ and $b=1,\; c=-2$ for the parameters in the electric charge generator in Eq.~(\ref{ch}) \cite{sap}. Quark family nonuniversality is present in these models because anomaly cancellation among the families requires us to have one family of quarks to transform differently from the other two under the gauge group.

Models based on the 3-4-1 symmetry predict the existence of two new neutral currents $Z^\prime$ and $Z^{\prime \prime}$ which mix up with the ordinary SM neutral current associated to the $Z$ boson. For models without exotic electric charges, the mixing can be constrained to occur between $Z$ and $Z^\prime$ only, so that $Z^{\prime \prime}\equiv Z_3$ is a mass eigenstate \cite{pgs,swz,spp}. Quark family nonuniversality generates two related effects: it leads to family-dependent left-handed couplings of quarks to the new neutral gauge bosons $Z^\prime$ and $Z^{\prime \prime}$ which are, in general, flavor nondiagonal thus leading to FCNC at low energy, and produces weak couplings of quarks to the neutral currents $Z_1$ and $Z_2$ (the mass eigenstates associated to the $Z-Z^{\prime}$ mixing) which depend on the family of quarks singled out as the one transforming differently.

Family dependence has been studied in this paper by identifying the three possible assignments of quark families in the weak basis into quark families in the mass basis. For the analysis we have selected two representative 3-4-1 models: Model A in the main text belongs to the class for which $b=c=1$ \cite{sap,pgs,swz}, and Model B belongs to the class for which $b=1,\; c=-2$ \cite{sap,spp,sen}. In Model A the $Z_3$ current couples only to exotic fermions, while the left-handed couplings of $Z^{\prime}$ to the SM quarks are flavor nondiagonal. In Model B, instead, ordinary quarks couple diagonally to $Z^{\prime}$, but nondiagonally to $Z_3$.

For Model A the three different assignments give three different allowed regions in the parameter space $\theta-M_{Z_2}$ (obtained from a fit to Z-pole observables and to APV data), thus producing different predictions for the lower bound on the $Z_2$ mass and for the range of values of the $(Z-Z^\prime)$ mixing angle $\theta$. These bounds have been further constrained by using experimental data from neutral meson mixing in the analysis of FCNC effects associated to quark family nonuniversality. For their study we have assumed the Fritzsch ansatz for the quark mixing matrix. The resulting new bounds show family dependence through the entries of the quarks mass matrices $V_L$ and $U_L$ to be replaced into the formulas for the mass differences in the $K^0-\bar{K}^0$, $B^0_d-\bar{B}^0_d$, $B^0_s-\bar{B}^0_s$ and $D^0-\bar{D}^0$ systems (see Eqs.~(\ref{mdifK}), (\ref{mdifB}) and (\ref{mdifD})). Combining both type of constraints leads to the conclusion that the heaviest family of quarks must transform differently in order to have a lower bound on $M_{Z_2}$ as small as possible. For Model A the smallest lower bound comes from the $B^0_d-\bar{B}^0_d$ system and turns out to be $M_{Z_2}>11.34$ TeV, which raises in 1 order of magnitude the lower bound $M_{Z_2}\geq 2.03$ TeV obtained from the allowed region in the parameter space $\theta-M_{Z_2}$. It must be, in any case, recognized that the bounds from neutral meson mixing are always obscured by the lack of knowledge of the mixing matrix entries and by the rather arbitrary assumed contribution of the $Z_2$ exchange to the mass differences. 

Model B has the particular feature that, notwithstanding two families of quarks transform differently under the $SU(4)_L$ subgroup, the three families have the same hypercharge $X$ with respect to the $U(1)_X$ subgroup. As a consequence, the couplings of the fermion fields to the neutral currents $Z_1$ and $Z_2$ are family universal. Thus, the allowed region in the parameter space $\theta-M_{Z_2}$ is the same for the three different assignments of quark symmetry eigenstates into quark mass eigenstates, and gives the family-independent bounds: $M_{Z_2}\geq 0.89$ TeV and $-0.00039 \leq \theta \leq 0.00139$. Since FCNC are present for this model in the left-handed couplings of ordinary quarks to the $Z_3$ gauge boson, the contribution of the $Z_3$ exchange to the mass differences in neutral meson mixing produces the family-dependent constraints on the $Z_3$ mass given in Eqs.~(\ref{lowerbB1}), (\ref{lowerbB2}) and (\ref{lowerbB3}). These results allows us to conclude that, like in Model A, the third family of quarks must transform differently in order to get, in this case, the smallest lower bound on $M_{Z_3}$ which comes from the $B^0_d-\bar{B}^0_d$ system and turns out to be $M_{Z_3}>6.61$ TeV. Since $M^2_{Z_3}=(g^2_4/2)(V^2$ $+v^2)$, this is also a lower bound on the scale of breaking of the 3-4-1 symmetry. As mentioned above, it must be stressed that the bound on $M_{Z_3}$ depends on inputs which are not dictated by the present experimental data, namely, the assumed ansatz for the quark mass matrices entries and the assumed contribution of the $Z_3$ exchange to the mass differences in the neutral meson mixing systems.

The convenience of distinguishing the heaviest family of quarks could give some indication as to why the top quark is unbalancingly heavy.

A comparison of the predictions from the two classes of 3-4-1 models without exotic electric charges shows that models for which $b=1,\; c=-2$ 
are preferred in the sense that they give lower bounds on the mass of the new neutral gauge bosons $Z_2$ and $Z_3$ smaller than the ones predicted by models in the $b=c=1$ class. In fact, first, the family-independent lower bound $M_2\geq 0.89$ TeV in Model B is not affected by the constraints coming from FCNC data and is just below the TeV scale and, second, providing the third family of quarks transforms differently, the lower bound on the $Z_3$ mass is $M_{Z_3}>6.61$ TeV, a value at the reach of the CERN LHC capability. This means that the $b=1,\; c=-2$ class of models have a better chance to be tested at the LHC facility or further at the ILC.  

Even though we have constrained ourselves to the particular case $V'\simeq V,\; v'\simeq v$, for which the mixing is present between the neutral gauge bosons $Z$ and $Z^{\prime}$ only, the analysis presented shows clearly the main purpose of this work, that is, the dependence of the predictions of 3-4-1 models on the possible quark family assignments associated to quark family nonuniversality. The extension to the general case of mixing between the three neutral currents present in the models would complicate the mathematical and numerical analysis, but the general conclusions regarding family-dependence of the phenomenology of 3-4-1 models would be the same.

Finally we note that, in order to make evident the effects of quark family nonuniversality, we have quoted only the lower bounds on $M_{Z_2}$ in Eqs.~(\ref{boundA1})-(\ref{boundsB}). As it is clear from Figs.~\ref{fig:1} and \ref{fig:2}, the $\chi^2$ fit also produces finite upper bounds on $M_{Z_2}$ which depend on the allowed value of the mixing angle $\theta$, except in the limit $\vert \theta \vert \rightarrow 0$ where $M_{Z_2}$ can be arbitrarily large. This is a characteristic feature both of 3-3-1 and 3-4-1 models without exotic electric charges associated to the fact that, if we consider the basic field content only, then the renormalization group equation analysis shows that, for all these models, the scale of gauge coupling unification $M_G$ can be as high as the Planck scale. This result can be modified in the 3-3-1 extension, for example, by considering $SU(6)$ as a covering group, and by introducing new physics at a scale $M_V \approx 2.0\; \mbox{TeV}< M_G \approx 3.0 \times 10^7\;\mbox{GeV}$ (where $M_V$ is the scale of the 3-3-1 symmetry  breaking) in the form of an enlargement of the scalar sector with an appropriate large number of Higgs fields that do not develop vacuum expectation values (for details see Ref.~\cite{gcu331}. See also Ref.~\cite{331gcu} for an alternative approach). This, clearly, prevents us to directly from getting an upper bound on the scale of the 3-4-1 symmetry breaking from the fit. Also, the constraints coming from FCNC processes allows us to put only lower bounds on the mass of the new neutral gauge bosons. Then, if we look only to these constraints, the 3-4-1 theory we have studied is not more predictive than generic grand unified theories where FCNC may also be suppressed by large masses. Notice, however, that bounds on other parameters can be obtained by examining additional phenomenological consequences of the model. For example, by taking into account that the 3-4-1 extension predicts new heavy particles and, provided these new particles feel the electroweak interaction, they should give corrections to electroweak precision measurements through their effects on the $W$ and $Z$ vacuum polarization amplitudes, which are usually expressed in terms of the oblique $S$, $T$ and $U$ parameters. It can be shown that in the general case $V'\neq V,\; v'\neq v$, the symmetry breaking pattern in Eq.~(\ref{break}) induces mass splitting between the new heavy gauge bosons different from $Z^\prime$ and $Z^{\prime\prime}$, and mass splitting between the extra heavy Higgs fields arising from the diagonalization of the scalar sector \cite{swz}, so these new particles may give no negligible contributions to the oblique parameters. A detailed study of these contributions, which we postpone to a future work, will enable us to put upper and lower bounds on their masses.

\section*{ACKNOWLEDGMENTS}
We acknowledge financial support from DIME at Universidad Nacional de Colombia-Sede Medell\'\i n.


\begin{thebibliography}{}
\bibitem[24]{correction}
Here we must point out that on account of an erroneous calculation of the term $\Delta Q^\prime_W$, the bound $0.67\;\mbox{TeV}\leq M_{Z_2} \leq 6.1\;\mbox{TeV}$ obtained in Ref.~\cite{spp} is incorrect. The correct calculation gives a fit which produces the lower bound $M_{Z_2}\geq 0.89$ TeV, with a finite upper bound depending on the allowed value of the mixing angle except for $\theta=0$, as shown in Fig.~\ref{fig:2}.
\end{thebibliography}
\end{document}